# Review of Electric Vehicle Charging Technologies, Configurations, and Architectures

**SITHARA S. G. ACHARIGE, (Graduate Student Member, IEEE),**
**MD ENAMUL HAQUE, (Senior Member, IEEE),**
**MOHAMMAD TAUFIQUL ARIF, (Member, IEEE),**
**NASSER HOSSEINZADEH, (Senior Member, IEEE)**
School of Engineering, Deakin University, VIC 3216, Australia

Corresponding author: Sithara S. G. Acharige (sgalwaduacharig@deakin.edu.au).

**ABSTRACT** Electric Vehicles (EVs) are projected to be one of the major contributors to energy transition in the global transportation due to their rapid expansion. The EVs will play a vital role in achieving a sustainable transportation system by reducing fossil fuel dependency and greenhouse gas (GHG) emissions. However, high level of EVs integration into the distribution grid has introduced many challenges for the power grid operation, safety, and network planning due to the increase in load demand, power quality impacts and power losses. An increasing fleet of electric mobility requires the advanced charging systems to enhance charging efficiency and utility grid support. Innovative EV charging technologies are obtaining much attention in recent research studies aimed at strengthening EV adoption while providing ancillary services. Therefore, analysis of the status of EV charging technologies is significant to accelerate EV adoption with advanced control strategies to discover a remedial solution for negative grid impacts, enhance desired charging efficiency and grid support. This paper presents a comprehensive review of the current deployment of EV charging systems, international standards, charging configurations, EV battery technologies, architecture of EV charging stations, and emerging technical challenges. The charging systems require a dedicated converter topology, a control strategy and international standards for charging and grid interconnection to ensure optimum operation and enhance grid support. An overview of different charging systems in terms of onboard and off-board chargers, AC-DC and DC-DC converter topologies, and AC and DC-based charging station architectures are evaluated. In addition, recent charging systems which are integrated with renewable energy sources are presented to identify the powertrain of modern charging stations. Finally, future trends and challenges in EV charging and grid integration are summarized for the future direction of EV charging systems researchers.

## I. INTRODUCTION

Electrification has become a major factor in social development, economic growth, and environmental contribution. Electrification is projected to increase further into the transport sector, focusing on the energy transition towards a zero-carbon emission economy. Electrified transportation is considered a desirable solution to reduce fossil fuel dependence and environmental impacts such as reducing greenhouse gas (GHG) emissions, climate change, and improving air quality. Electrified mobility will open the possibility of using alternative energy systems such as renewable energy sources (RESs) to secure mobility and make road transport more independent from fossil fuels. Electric vehicles (EVs) offer zero-emission, highly reliable, efficient, and low-maintenance vehicles compared to conventional internal combustion engine (ICE) vehicles. The advanced power controllers and smart charging technologies in modern EV charging systems can control optimum charging and discharging, and dynamic power sharing by communicating with EVs and the utility grid to improve the energy efficiency of chargers and decrease pressure on the local power grid.

The deployment of EVs will depend on a driving range, model, performance, costs of batteries, the convenience of re-charging, safety perception and possible implied driving habits [1]. The charging time needs to be matched with the time required for conventional ICE vehicles, and EV supply equipment (EVSE) systems need to extend with higher power levels for ultra-fast charging. Smart charging coordinated control techniques, and high-power converters can be used to reduce the charging time. The EV battery is playing a significant role in the charging system for supplying energy to EVs as well as acting as energy storage for the utility grid. The power flow of the EV charging system can be unidirectional or bidirectional. Most of the commercial onboard chargers are equipped with unidirectional power flow, and grid-to-vehicle (G2V) capabilities due to simplicity, reliability, low cost, and simple control strategy. In contrast, bidirectional chargers can inject power into the utility grid through vehicle-to-grid (V2G) other than G2V operation. Hence, bidirectional chargers are considered active distributed resources with specific control modes to support load leveling, and RES integration, prevent overloading EV loads, and reduce power losses in the utility grid. Therefore, researchers are becoming more interested in bidirectional chargers as a potential option for EVs in the future.

However, high EV integration into the distribution grid can create challenges and concerns such as increased peak demand, increasing harmonic distortions, frequency, and voltage deviations, power quality impacts, high active power absorption, and unpredictable dynamic behaviors. Moreover, EVs require a considerable amount of electric energy through the power grid, which is mostly delivered by the power grid from fossil fuel-based power plants. Thus, EVs may contribute to carbon emissions depending on the source of energy used to charge the battery. The increased EV charging station may change the distribution network's load patterns, characteristics, and safety requirements. Government policies and standards for EVs have been introduced internationally to ensure reliable grid operation and mitigate negative impacts on the distribution grid [2]. Therefore, many research studies have established advanced control strategies and converter topologies for EV charging systems.

The RESs integrated EV charging systems have gained interest in the industry as a cost-effective, clean, and sustainable technique to charge EV batteries. RESs are capable to provide ancillary services to the power grid by reducing peak demand and improving power quality, energy efficiency, and reliability. Various EV charging systems have been introduced with solar PV, wind power, energy storage systems (ESSs), supper capacitors, and fuel cell to enable low emission, highly flexible, and economical power systems [3],[4],[5]. Among different RESs, solar PV powered EV charging stations are widely established due to their technological advancement. The ESSs are becoming an integral part of EV charging systems along with the RESs in microgrid and smart grid frameworks. The authors in s proposed a hybrid optimization algorithm for ESS and solar PV integrated EV charging stations to reduce grid EV charging cost. In [7], a grid-based EV charging system is designed with multiple sources including solar PV, ESS, and diesel generator to provide constant charging in grid connected and islanded modes. The decentralized EV charging optimization technic for building integrated wind energy is presented in [8] with real-time coordination.

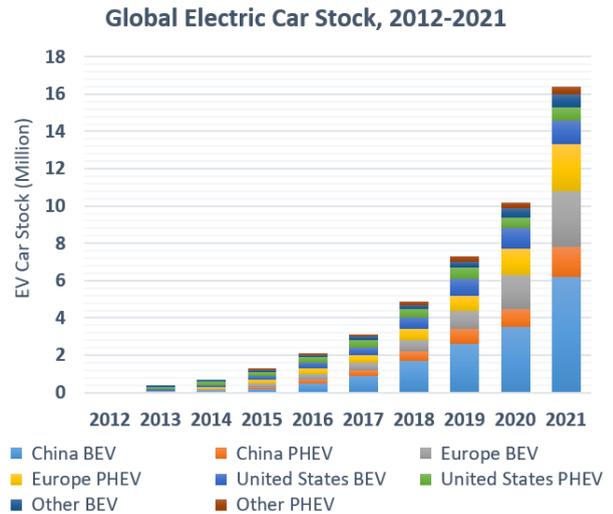

**FIGURE 1.** Global electric passenger car stock, 2010-2021 [6]

The charging system is required a dedicated converter topology, and control technique and is necessary to follow a set of standards to integrate with the utility grid. Recently, many international standards and codes regarding EV charging and utility interface have been introduced to achieve widespread EV acceptance and reliable grid operation. The international organizations are established various standards and codes, universal structures, associate peripheral devices, and user-friendly software for EV charging systems toward the successful growth of EVs on the power network. Fast-charging standards for AC and DC charging have recently improved significantly by IEEE, IEC, and SAE organizations [9]. The EV charging systems are comprised of several AC/DC power converters and advanced control algorithms to operate G2V and V2G operations while providing grid support. The AC-DC and DC-DC converters are employed to deliver power either unidirectional or bidirectional in charging systems. The cost, size, performance, and efficiency of the charging system is depending on the corresponding converter topology.

Therefore, a comprehensive review of AC-DC and DC-DC converters is important to identify prevailing challenges and find remedial solutions. Most EV charging stations are AC power-based configurations as they have matured technology and direct usage of local loads when compared to the DC power based fast charging stations. On the other hand, DC power-based architectures are becoming popular in recent years due to their high efficiency, low cost, and flexibility to integrate RESs and ESS in the distribution grid. However, complexity is increased with additional energy sources in DC power-based charging stations. Hence, EV charging topologies, configurations, and architectures need to be analyzed comprehensively to identify the current states of EV charging systems, technical development, and challenges to identify a remedial solution. The following contributions are made in this paper.

- Current state of EV charging technologies, international standards, and codes are reviewed based on charging levels, modes, types of connectors and ports, and corresponding EV standards and grid codes.
- Onboard and offboard EV charging technologies are comprehensively analyzed with associated powertrain.
- EV charging configurations and converter topologies including AC-DC converters and DC-DC converters are reviewed
- Finally, EV battery technology, architectures of EV charging stations, and future trend and challengers are discussed.

The rest of the paper is organized as follows. The current state of EV charging technologies is presented in section 2. In section 3, overview of EV technologies and charging requirements such as different types of charging levels, modes, and connectors are presented. The international EV standards including charging, grid integrated, and safety are reviewed in section 4. EV charging topologies are comprehensively reviewed based on power flow, G2V, and V2G operation, onboards and offboard chargers in section 5. Configurations and converter topologies of EV chargers are analyzed in section 6 based on the AC-DC and DC-DC converters. The EV battery technology and different EV architecture of charging stations are reviewed in sections 7 and 8 respectively. Finally, future trends and challenges of EV charging are discussed in section 9, and the conclusion is drawn in section 10.

## II. CURRENT STATE OF EV CHARGING TECHNOLOGY

Electrified transportation is achieving momentum in the current industry due to many factors, including clean environmental concepts, fossil fuel depletion, affordability, increasing charging infrastructures, and smart control propulsion strategies. The EV market is expanding rapidly and over five times as electric car models available in 2021 as compared to 2015. In 2021, global EV sales doubled from the previous year to a record of 6.6 million. The global electric car sales were 2 million in the first quarter of 2022, 75% of the same period in 2021. The projections indicate that the global EV fleet will reach 230 million vehicles in 2030 and 58% of vehicles are expected to be electric vehicles in 2040 [10],[11]. Global electric car stock is significantly increased in 2021 when compared to the previous years and the total number of electric cars on road to over 16.5 million shown in Fig. 1. The largest EV market belongs to China which contributed 50% of the EV growth in 2021 and was followed by Europe and the United States. Electrified transportation has attracted much attention from governments and private stakeholders to move towards carbon neutrality in 2040 through consistent policy support, incentives, and subsidies from the governments.

Several major automakers are preparing for a shift to EVs and some countries, including Europe and China are planning to restrict fossil fuel-powered vehicle sales in the future [12]. The exponential growth of EV sales will be expected instantly as increasing high charging equipment accessibility, public fast charging stations, and the development of RESs. Most of the projects and developments are focusing on the RESs powers changing infrastructures to make further advantages from EVs while supporting grid operations [13]. As a result, carbon footprint declines, and additional grid support is provided by EVs through charging systems. The adoption of EVs is the consequence of many regulatory assistances, such as purchase incentives, subsides for home charging insulation, benefits for drivers and parking, introduced international standards, and expanded accessibility to public charging infrastructure [14].

Battery technology has a significant impact on the growth of EVs because the price, weight, volume,charging time, and lifetime of the battery are significant hurdles to commercialization which are the main research topics in the industry. The cost of EV batteries has decreased significantly from over $ 1000/kWh in 2010 to about $132 in 2021. Most analysts predict that the cost of battery packs will keep declining, reaching $100/kWh between 2023 and 2025 and $61–72/kWh by 2030 [15]. The investments in EVs have drastically increased to explore electrification strategies to increase driving range, efficiency, and reliable charging and discharging capabilities for an affordable price. The inductive and wireless charging systems are interesting topics in current EV technology which is still at an early stage. The power converter topologies are responsible for the efficient and reliable performance of fast EV charging systems. Another attractive research area is the V2G application, EV batteries can be used to store surplus energy and feedback energy to the utility grid using coordinated control strategies [16]. Moreover, the widespread availability of fast-charging stations will start a movement where electric vehicle charging will become as common as refueling ICE vehicles at existing service stations.

Technological innovations in EVs have provided new concepts for EV grid integration, offering attractive and competitive regulated charging/discharging techniques. [17]. The EVs will play a new role in the emerging concept of smart charging technologies by exchanging energy with microgrids and the power grid via a bidirectional power flow with ancillary support [18]. The smart EV fleets programs encourage to design of integrated EVs into their transportation system via RESs [19]. However, the rapid growth in EVs is expected to have further negative impacts on the distribution grid including power quality impacts, increasing peak demand, voltage instability, harmonic distortion, and overloading distribution grid. Therefore, extensive research studies have been progressing to identify the power network and environmental and economic impacts of EVs [20].

## III. ELECTRIC VEHICLES AND CHARGING REQUIREMENTS

### A. TYPES OF ELECTRIC VEHICLE

The electric vehicle comprises one or more electric motors, controllers, and a charging system with a battery pack for energy storage. The electric motor either assists completely via electric power or ICE depending on the EV type. The battery pack is mainly recharged from electric energy through a charging system. Additionally, the electric motor functions as a generator and provides power to charge the battery using a bidirectional DC-AC converter during the braking and deceleration of the vehicle. Conversely, the converter enables power to flow from the battery to the motor during driving mode [21]. Based on the current phase of development, EVs are categorized into two types hybrid vehicles and all-electric vehicles (AEVs) by considering the degree of use of electricity as shown in Fig. 2.

The hybrid vehicle has a conventional ICE vehicle design and a battery to power the vehicle using fuel and electric energy. The capacity of the battery defines the driving range of the vehicle in electric mode. Hybrid electric vehicles (HEVs) and plug-in hybrid vehicles (PHEVs) are two types of hybrid vehicles in the market. The hybrid vehicle comprises ICE, an oversized electric motor, and a battery to reduce fossil fuel consumption. The HEVs have similar drive features to normal ICE vehicles and can make the vehicle move using stored battery power for short distances. The battery pack automatically chargers through regenerative braking by turning kinetic energy into electric energy when ICE is at a light load in HEVs [22],[23]. The electric motor can contribute to and improve the efficiency and performance of the hybrid vehicle when the ICE is under a heavy load. The propulsion mechanism of the PHEV is like the HEV, and it differs from the HEV by having a large battery pack being able to charger from the regenerative breaking and plug into the power grid. The PHEV has a more powerful electric motor than HEV which is enabled to be operated in entirely electric mode by turning off the ICE [24],[25]. The all-electric driving range is about 25 km to 80+ km in PHEV depending on the model.

AEV uses electric power as fuel to recharge the battery pack and consists of electric motors for propulsion. AEVs produce zero pipeline emissions as they are solely driven via electric power without any fuel combustion. There are three types of AEVs including battery EV (BEV), fuel cell EV (FCEVs and extended- range EV (ER-EV). The AEVs use a large onboard battery pack to provide acceptable propulsion to the vehicle without any form of ICE. BEVs are frequently called EVs which are solely driven by electric motors powered via a battery pack. The BEV is exclusively powered by electricity and thus tends to have a large battery capacity (kWh) when compared to hybrid vehicles as they are relying only on electric power. The battery pack chargers by plugging the vehicle into the power grid or electric source and regenerative braking. The main challenges of BEV are shorter driving range per charge, limitation of public charging stations, and long charging period [26],[27]. The modern BEVs are designed to have a driving range of 160 km to 650+ km depending on the vehicle model. Moreover, the deployment of BEVs can support the power grid via smart charging technologies and V2G to increase variable renewable energy and interplay with communication technologies to minimize operational costs and maximize the technical features of the power systems.

The fuel cell electric vehicles are powered by hydrogen gas and the propulsion system is the same as EVs. The FCEV uses hydrogen gas to power an electric motor entirely by electricity that is not required fuel or changing from the grid. The FCEV is refueled with hydrogen gas and fuel cell uses to transform the chemical power into electric energy which drives the electric motor. The FCEV has a short refueling time and the driving range is comparable to ICE vehicles. Moreover, fuel cell cars are noise-free, very energy efficient, and have zero tailpipe emission vehicles which produced pure water as a waste [23]. Extended-range electric vehicles comprise an electric motor and small ICE to produce additional power which is used to keep the battery charging for long distances. ER-EVs are AEVs with many of the benefits of purely electric models. The E-REVs help combat range anxiety, lower fuel costs, and are highly efficient and maximize the use of their vehicles by operating their engine constantly [28]. An ER-EV comprises with electrical drivetrain (one or more electric motors and battery pack) and ICE to charge the battery pack. The range-extender vehicle is only driven by the electric motor and the sole purpose of ICE is to recharge the battery pack of the EV. Therefore, ER-EV needs to be both recharged from the power grid and refueled at a petrol station. The specifications of different type of popular EVs are presented in Table 1.

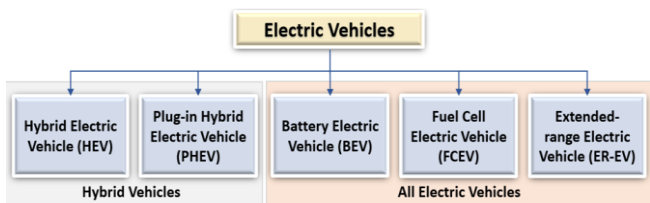

FIGURE 2. Types of electric vehicles

TABLE I
SPECIFICATIONS OF COMMERCIAL ELECTRIC VEHICLES [29], [30], [31]

| VEHICLE MODEL | TYPE | BATTERY CAPACITY (kWh) | DRIVING RANGE (km) | Connector Type |
|---|---|---|---|---|
| Chevrolet Volt | PHEV | 18.4 | 85-Battery | Type 1 J1772 |
| Mitsubishi Outlander | PHEV | 20 | 84-Battery | CCS, Type 2 |
| Volvo 40 series | PHEV | 78 | 410 | CCS, Type 2 |
| Nissan Leaf Plus | BEV | 64 | 480 | CHAdeMO, Type 2 |
| Tesla Model S | BEV | 100 | 620 | Supercharger |
| Tesla Model X | BEV | 100 | 500 | Supercharger |
| Tesla Model 3 | BEV | 82 | 580 | Supercharger |
| Kia Niro- SUV | BEV | 64 | 460 | CCS, Type 2 |
| Lexus UX 300e | BEV | 54.3 | 320 | CHAdeMO, Type 2 |
| Ford Mustang | BEV | 70 | 400 | CCS, Type 2 |
| Jaguar ev400 | BEV | 90 | 450 | CCS, Type 2 |
| Renault Zoe | BEV | 52 | 390 | CCS, Type 2 |
| BMW i3 | BEV | 37.9 | 310 | CCS, Type 2 |
| Toyota Prius Prime | PHEV | 8.8 | 40-Battery | SAE J1772 |
| Honda e | BEV | 28.5 | 220 | CCS, Type 2 |
| Porsche Taycan | BEV | 93 | 410 | CCS, Type 2 |
| Volkswagen e-Golf | BEV | 35.8 | 230 | CCS, Type 2 |
| Mercedes-EQA | BEV | 66.5 | 420 | CCS, Type 2 |
| Audi e-tron | BEV | 95 | 400 | CCS, Type 2 |
| BMW iX3 | BEV | 80 | 460 | CCS, Type 2 |
| Toyota Mirai | FCEV | 1.6 | 502 | - |
| Hyundai Nexo | FCEV | 40 | 570 | - |
| Honda Clarity | FCEV | 25.5 | 550 | - |
| BYD Atto 3 | ER-EV | 60.4 | 420-Battery | CCS, Type 2 |
| Hyundai Kona | ER-EV | 64 | 577-Battery | CCS, Type 2 |

## B. CHARGING LEVELS AND MODES

EVs are designed with various charging technologies, capacities, charging and discharging strategies to fulfill their unique requirements. Therefore, standardized charging levels and models are established to drive EV adoption forward in the industry and help to enable innovative developments, and research studies with general acceptance. The electric powertrain of modern plug-in electric vehicles is similar and is designed with a high-power battery pack to maintain voltage and current, a battery management system, various converters to supply appropriate voltage levels, controllers and drive inverters [32], [33]. EV chargers can be classified as onboard and offboard chargers as well as unidirectional and bidirectional EV chargers. Charging technologies can be classified as conductive charging, battery swapping, and wireless charging as shown in Fig. 3. The slow and fast charging systems are integrated with a low-voltage distribution grid. Extreme fast chargers (XFC) are designed the present with a medium voltage power grid to replicate the existing fuel station infrastructure. Three types of charging levels and four charging modes are established based on the power level and safe communication constraints respectively.

### 1) CHARGING LEVELS

Conductive charging involved an electric connection between the charging inlet and the vehicle which follows three charging levels such as Level 1, Level 2, and Level 3 depending on the power level as shown in table 2. Level 1 and level 2 charging is used in onboard chargers with AC power while level 3 charging is used in offboard chargers with DC power. Level-1 and level-2 charging EV service equipment (EVSE) must be on-board and level-3 EVSE must be off-board chargers and following the SAE J1772 standard. The level 1 charger uses a 120 V single-phase AC power supply and has the slowest charging speed which is generally used in domestic with low power level (up to 1.92 kW) without any additional infrastructure [34]. The conductive connection of level 1 uses a SAEJ1772 standard connector and the lowest charging rate can be used in off-peak time [35]. Therefore, level 1 charging is especially appropriate for long-time or overnight charging. Depending on the battery capacity and type, charging level 1 generally required about 3-20 hours to fully recharge the battery pack and standard outlets are available in all places.

Level 2 charging is the main charging method of private and public facilities which is compatible with both PHEV and BEV and. Level 2 charging able to provide power up to 20kW under 208Vac or 240 Vac input voltage with a maximum 80 A current. Level 2 chargers are equipped with onboard chargers and required dedicated components and installation. Level 2 charging connector use IEC62196-2 Type-2 standard in Europe, SAEJ1772 Type 1 or Tesla super chargers in US [36],[37]. EV owners are interested in level-2 chargers because of their fast-changing rates (4 - 6 hours) compared to the level 1 chargers and standardized charger-to-vehicle connections. DC fast charging or Level 3 charging uses 480 Vac electric power to deliver high voltage DC power to the EV battery. Level 2 chargers can handle high power rage in between 50 kW to 350 kW to supply dc voltage around 300 V to 800 V in offboard chargers within 30 minutes. The volume and the weight of the chargers are reduced as level 2 chargers are placed outside of the vehicle due to high power flow. DC fast chargers are directly connected to the vehicle via offboard chargers and CHAdeMO, Tesla superchargers and CCS combo 1,3 connectors are considered for level 3 fast charging. The charging time of the level 3 charging is less than 1 hour and reduces the driving range anxiety of EV owners.

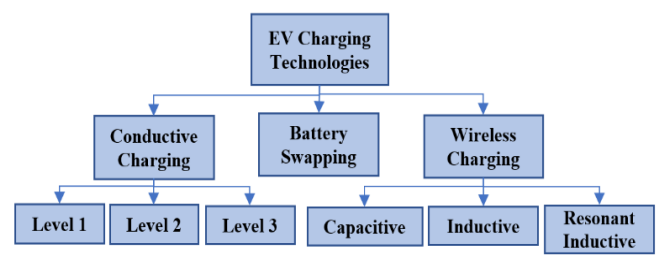

**FIGURE 3.** Electric Vehicle Charging Technologies

TABLE II
COMPARISON OF DIFFERENT ELECTRIC VEHICLE CHARGING LEVELS [34]

| SPECIFICATION | LEVEL 1 | LEVEL 2 | LEVEL 3 | EXTREME FAST CHARGING (XFC) |
|---|---|---|---|---|
| **Charging Power** | 1.44kW - 1.9kW | 3.1kW - 19.2kW | 20kW – 350kW | >350 kW |
| **Charger Type** | Onboard - Slow Charging | Onboard - Semi-Fast Charging | Offboard Charger Fast Charging | Offboard Charger Ultra-Fast Charging |
| **Charge Location** | Residential | Private and Commercial | Commercial | Commercial |
| **Charging time** | 200km: +/- 20 hours | 200km: +/- 5 hours | 80% of 200km: +/- 30 min | Approximately 5 min with high energy density |
| **Power Supply** | 120/230Vac, 12A - 16A, Single-Phase | 208/240Vac, 12A - 80A, Single Phase/split phase | 300-800Vdc, 250-500A Three-Phase | 1000Vdc and above, 400A and higher - Polyphase |
| **Supply Interface and Protection Type** | Convenience outlet (Breaker in cable) | Dedicated EV Supply Equipment (Breaker in the cable and pilot function) | Dedicated EV supply equipment (communication & event monitoring between EV and charging station) | Dedicated EV supply equipment (communication & event monitoring between EV and charging station) |
| **Standards** | SAE J1772,  IEC 62196-2, IEC 61851-22/23, GB/T 20234-2 | | IEC 61851-23/24 IEC 62196-3 | IEC 62196 SAE J2836/2 & J2847/2 |

However, low power chargers, level 1 and level 2 have the lowest negative impact on the power network during peak time. The local distribution grid may become overloaded by the level 3 chargers due to high power usage during peak times [38]. Extrema fast charging (XFC) systems have the capability to deliver a refueling experience like ICE vehicles. The XFC systems can handle more than 350 kW power with 800 Vdc internal DC bus voltage and battery recharging time is approximately 5 min. The XFC stations are designed with power electronic components focusing on solid-state transformers (SST), isolated DC-DC converters, and front-end AC-DC converter stages and controllers. The installation cost of the XFC is very high and required dedicated EVSE to deliver high power. The XFC station can be designed by combining several XFC systems to provide a chance to lower operational and capital investment to make it economically feasible. Additionally, SST provides advantages over conventional line-frequency transformers for converting medium voltage levels into low voltage levels and providing galvanic isolation in XFC stations [39].

### 2) CHARGING MODES

The International Electrotechnical Commission (IEC) defines 4 charging modes (IEC-62196 and 61851) for AC and DC charging systems and provides the general attributes of the safe charging process and energy supplied requirements [40]. The slow charging applications follow mode 1 which comprises with earthing system and circuit breaker for protection against leakage and overloading conditions. The current limit of mode 1 varies from 8 A to 16 A depending on the country. EV is directly connected to the AC grid either 480 V in three-phase or 240 V in single-phase via a regular socket in mode 1. The cables are integrated with a specific EV protection arrangement in mode 2 slow charging from a regular power outlet to supply moderate power. Mode 2 has a direct and semi-active connection of the AC grid to EV with a maximum 32A current flow. The mode 2 cable provides our-current, overheat protection, and protective earth detection.  Therefore, mode 2 charging cables are more expensive than mode 1 due to high current flow and provide minimum standards and moderate safely for modern EVs [41].

Mode 3 is used for slow or fast charging via a specific outlet with the controller. The dedicated circuit is permanently installed for protection and control in this mode to meet standards in domestic or public charging stations. Mode 2 allows a higher power level with a maximum 250 A current which is used fixed EVSE for single-phase or three-phase grid integration. The connection cable includes an earth and control pilot to enable proper communication between the EV and the utility grid. The fast-charging station uses mode 4 via fixed EVSE such as CHAdeMO charging technology with 600 V DC power. Mode 4 chargers are more expensive than mode 3 and the connection includes earth and control pilot to control a maximum 400 A current. Offboard chargers are following mode 4 specifications with a wide range of charging capabilities over 150 kW power [42]. The summery of the charging modes are shown in table 3.

### C. CHARGING PORTS AND CONNECTORS

EV charge components including power outlets, connectors, cords, and attached plugs are the main components of EVSE which provide reliable charging/discharging of EV battery and protection. The configuration of charging peripheral devices, power ratings, and standards are varied from location and country. Commercially available different ports and connectors are shown in table 4 and 5 by following their standards respectively. However, manufacturers are attempting to introduce common charging EVSE to avoid standards conflicts and difficulties [44]. Both AC and DC charging systems followed SAE J1772 standards connectors in the U.S and IEC 62196 standards in Europe. In Japan, AC charging ports and connectors are followed SAE J1772 standards, and DC fast charging ECSE are relying on the CHAdeMO. Most of the DC charging stations are employed with the CHAdeMO protocol due to technological advancements which are developed by Japanese manufactures [45]. The Chinese automaker has developed and used their own DC fast charging protocol known as GB/T 20234, or simply GB/T DC fast charging. Different EV charging ports and connectors of Ac and DC charging with their standard are shown in Tables 4 and 5.

TABLE III
COMPARISON OF DIFFERENT CHARGING MODES [42], [43]

| CHARGE MODE | PHASE | CURRENT | VOLTAGE | POWER (MAX) | SPECIFIC CONNECTOR | CHARGING CONFIGURATION |
|---|---|---|---|---|---|---|
| **Mode 1** | AC - 1Φ AC - 3Φ | 16A | 230-250V 480V | 3.8 kW 7.6 kW | No | 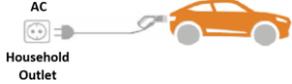 |
| **Mode 2** | AC - 1Φ AC -3 Φ | 32A | 230-250V 480V | 7.6 kW 15.3 kW | No | 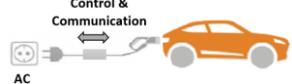 |
| **Mode 3** | AC - 1 Φ AC - 3 Φ | 32-250A | 230-250V 480V | 60 kW 120 kW | Yes | 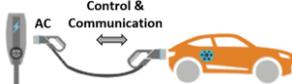 |
| **Mode 4** | DC | 250-400A | 600-1000V | >150kW | Yes | 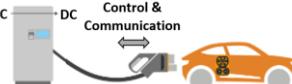 |

The Australian standard for EV charging plugs and connectors (IEC 62196) encourages the adoption of both US and EU connector standards rather than imposing a single standard [46]. The IEC 62196 standard specifies four types of coupler configurations for DC fast chargers. They are configuration AA (CHAdeMO), configuration BB (GB/T), configuration EE (CCS-Type 1) and configuration FF (CCS-Type 1) [47]. Tesla uses their connector (superchargers) with the ability to support both AC charging and DC fast charging systems. Tesla superchargers offer excellent fast charging speeds via their own designed charging stations and connectors can supply 72 kW, 150 kW, or 250 kW electric power [48]. The Combo connectors and ports are another popular DC fast charging component used in Europe which has the same appearance as AC J1772 and additional DC charging pins in the same AC charging connector with quite a different shape. Governing bodies and manufacturers are attempting to ensure compatibility by developing standardized standards, protocols, and couplers for fast and ultra-fast charging systems at present.

## IV. STANDARDS OF ELECTRIC VEHICLE CHARGING AND GRID INTEGRATION

Standards play a key role in the deployment and development of EV technology in society which serve as a crucial foundation for broad market penetration and customer satisfaction. The high level of EV charging integration has created new challenges and requirements in the automotive industry and electric networks. Standards and grid codes are designed to ensure reliable and safe EV integration with the power grid and other energy resources. Charging standards are applied to EVs to provide accurate functionality, protection, interoperability, and integration with various parameters and conditions [49],[50]. Many EV charging standards are employed around the world to interact with charging infrastructure. The Society of Automotive Engineers (SAE) and the Institute of Electrical and Electronics Engineers (IEEE) are two main contributors to charging and grid integration standardizations. The SAE and International Electrotechnical Commission (IEC) standards are widely used for EV conductive charging systems.

Charging standards and regulations can be categorized as charging components, grid integration, and safety [51]. The specifications of EV conductive charging components including connectors, plugs, outlet-socket, and inlets are mainly provided by SAE J1772 and IEC 62196 standards. series of standards in IEC 62196 and IEC 61851 provided the specification for EV connectors in Ac and DC charging systems. Inductive charging standards are SAE J1772, IEC 61980, and battery swapping charging systems used IEC 62840 standards [52]. AC charging systems comprises SAE J1772 standards with 100V domestic power in US and Japan and 220V power in Europe. GB/T 20234 standards are employed in AC charging systems in China. Connectors and ports in DC charging systems are designed by using a set of IEC 61851 standards, CHAdeMO which is described in GB/T 20234, and CCS Combo standards [54],[55].

TABLE IV
DIFFERENT PORTS OF AC AND DC FAST CHARGING [53], [11]

| Charging Type | Japan | USA | | Europe | | | China | ALL Markets |
|---|---|---|---|---|---|---|---|---|
| AC Charging | 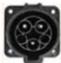 SAE J1772 IEC 62199/6, IEC 61851-22/23 Level 1,2 1Φ, split Phase | | | 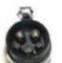 IEC 62196 Level 1 1Φ | 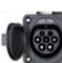 IEC 62196-2 Level 2, 3 1Φ and 3Φ | 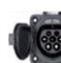 IEC 62196 Level 1,2 1Φ and 3Φ | | 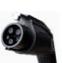 Tesla IEC 62196-2 1Φ |
| DC Fast Charging | 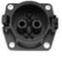 JEVS G105 CHAdeMO | 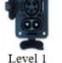 Level 1 DC SAE J17772 -2017 IEC 61851-23/24 IEC 62196-3 | 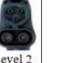 Level 2 DC | 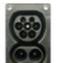 IEC 62196-3 IEC 61851-23/24 Hybrid Combo | | | 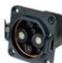 GB/T 20234-3-2015 IEC 62196-3 | 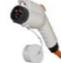 Tesla IEC 62196-3 1Φ |

TABLE V
DIFFERENT CONNECTORS OF AC AND DC FAST CHARGING

| Charging Type | Japan | USA | EU | China |
|---|---|---|---|---|
| AC Charging | 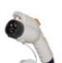 SAE J1772 Level 1, 2 1Φ and 3Φ | 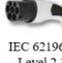 SAE J1772- Level 1, 2 1Φ and 3Φ | 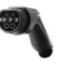 IEC 62196-2 Level 2, 3 1Φ and 3Φ | 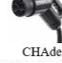 IEC 62196 Level 1,2 1Φ and 3Φ |
| DC Fast Charging | 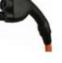 CHAdeMO IEC 61851-23/4 IEC 62196-3 Globally accepted | 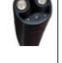 SAE J1772 Level 2- DC | Tesla Super Charger | 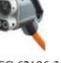 IEC 62196-3 Hybrid Combo GB/T 20234-3 |

Charging and discharging of EVs through the grid is controlled by grid integration standards and codes. The EV is considered as a distributed energy resource in V2G operation mode which is applied power grid integration EV standards. Grid integration standards include power regulations, safety, and power quality requirements, and important grid codes to ensure reliable integration of EVs. Grid interconnection standards and regulations are generally established by the Institute of Electrical and Electronics Engineers (IEEE), and Underwriters' Laboratories (UL) organizations. Standards for interconnection of distributed resource in the power grid is included in IEEE 1547 which explains the performance, maintenance, testing, and safety requirements of all DER on distribution systems [56]. Power converters, controllers, and safety specifications of DERs are presented in UL 1741 standards [57]. Communication standards in IEEE 2030.5 and ISO 15118 provide interoperable control for EVs via information exchange, test procedures, response specification, and security requirements [58].

The EVSEs are used to communicate with the EV to ensure a safe and appropriate power supply other than delivering energy in between the EV battery and energy source. Therefore, some standards are developed for signaling and communication with multiple devices. The primary objective of communication standards is to regulate the amount of current provided and manage the current flow of different devices. Moreover, the state of charge (SoC) of the battery also monitors and allows to use of EVSEs. Communication specifications of DC offboard fast charger is designed with SAE J2847/2 standards [52] and PLC communication requirements can be observed in SAE J2931/4. International Organization for Standardization (ISO) is also developed many safety-related standards and technical regulations for lithium-ion battery packs (ISO 6469 1-3) and EVs in high voltage systems (ISO/DIS 21498) [59]. Table 6 lists the genatrally preferred international standards for EV charging systems including conductive charging, safety, and grid integration regulations. Table 7 defines the globally established standards, which supervise different characteristics of EVs.

## V. ELECTRIC VEHICLE CHARGING TOPOLOGIES

The expanding popularity of EVs results in various types of charging topologies, control strategies, converters, power requirements, and charging stations to maximize energy efficiency while satisfying the constraint of both EVs and the utility grid. Several articles have been summarized EV structure and charging configuration [56],[60],[61]. Modern PEVs share a similar powertrain, which is comprised of a high voltage battery pack to sustain moderate currents, onboard charger, battery management system, drive inverters, DC-DC converters, and high voltage loads such as cooling system, heaters [11]. EVs are highly depending on the energy storage technique including high voltage battery packs, supercapacitors, and fuel cells. Therefore, charging technology provides an essential link between the EV and energy supply resources. BEVs can be charged from AC and DC power via EVSE by communicating with the EV and the charger to ensure an efficient and safe energy supply [52]. For EV charging systems, three charging methods are employed including conductive, inductive charging, and battery swapping as shown in Fig. 4 [62]. Both conductive and inductive or wireless charging have advantages over each other in terms of convenience, reliability, and efficiency, thus it is anticipated that both types of chargers will exist at the same time in the future EV market [63], [64].

TABLE VI
MAIN INTERNATIONAL STANDARDS FOR EV CHARGING SYSTEMS

| STANDARD | DESCRIPTION |
|---|---|
| SAE J1772 | Conductive charger coupling of AEVs and HEV |
| SAE J2344 | Guidelines for EV safety |
| SAE J2894/2 | Power quality requirements |
| SAE J2953 | Standards for interoperability of EV and charger |
| SAE J2847/1 | Communication between EV and the grid |
| SAE J3068 | EV power transfer system using a three-phase AC capable coupling |
| SAE J2931/7 | Security for PEV communication |
| IEC 60038 | Standards for the voltage for charging applications |
| IEC 62196 | Standards for EV conductive charging components (outlets, plugs, connectors, and inlets) |
| IEC 60664-1 | Installation coordination for charging equipment in low voltage supply |
| IEC 62752 | Standards for cable control and protection devices |
| IEC 61851 | Covering safety-related specifications on the charging station |
| ISO 15118 | Standards for V2G communication protocols and interfaces |
| ISO 17409 | Specifications for the connection of EV with an external energy source |

A comprehensive review of conductive charging technology is presented in this paper. Many automakers equip their vehicles with both AC and DC chargers, giving customers great flexibility in charging their vehicles at home or in public charging stations [66]. Level 1 or level 2 chargers are designed used for charging EVs at home, whereas level 2 and level 3 or DC fast chargers are found at public charging stations [67]. Most EV chargers are compatible with a wide range of EV models [52]. EV manufactures include both AC and DC chargers in the same vehicle to enable either onboard or offboard charging capabilities as shown in Fig. 5. Furthermore, EV battery chargers have various AC and DC power converters to provide high efficiency, reliability, and power density and either through coordinated or uncoordinated control [68]. EV chargers use either AC or DC power supplies to recharge the battery pack with specific power ratings, standards, and components. AC charging is the common method used in EVs which is converted AC-DC inside the EV in the onboard charger and then charge the battery [69]. The charging speeds depend on the converter capability and output power level of the charging point. Conventional AC chargers have limited power, less than 22kW, and required longer charging time. The DC charging used in fast chargers uses an off-board circuit to generate a high voltage (300 -1000V) [70].

The DC chargers convert power before entering the EV in the dedicated offboard charger and then directly charger the battery from DC power bypassing the in-built converter inside the vehicle. DC charging requires high power (20 kW - 350 kW), specific components, safety protocols, and large power control circuits to control high power levels. EV charging control systems can be classified as uncoordinated and coordinated or smart chargers. The battery starts to recharge instantly when plugged in or after a user-fixed delay in uncoordinated charging systems [71]. Therefore, uncoordinated chargers can cause a significant impact on the power grid when unpredictable EV charging loads arrive and lead to high peak demand over loading, and power quality impacts [72]. Hence, well-synchronized charging coordination between EVs and grid operators is essential to maximize the load factor and minimize the power losses while enabling grid support [73]. EV charging configurations such as G2V and V2G, and onboard and offboard charging systems are explained in this section.

TABLE VII
INTERNATIONAL EV CHARGING STANDARDS AND GRID CODES

| ORGANIZATION | STANDARDS | DESCRIPTION |
|---|---|---|
| The Institute of Electrical and Electronics Engineers (IEEE) [17] | IEEE 519-1992 | Harmonic control in electrical power system |
| | IEEE 1159-1995 | Monitoring electric power quality |
| | IEEE 1100-1999 | Powering and grounding sensitive electronic equipment |
| | IEEE 1366-2012 | Electric power distribution reliability indices |
| | IEEE1547 | Standards for interconnecting distributed resources with electric power systems (10MVA or less PCC) |
| | P1547, P2100.1 | Standards of different aspects of grid connection of DERs, charging system standardization |
| Society of Automotive Engineers (SAE, United States) [22] | SAEJ2293 | - Standards for on-board and off-board charging equipment (Conductive AC and DC, inductive charging)<br>- Power requirement, system architecture for conductive AC, DC, and inductive charging<br>- Communication and network requirements of EV charging [53] |
| | SAEJ1772 | Ratings for all the equipment for EV charging- (voltage and current ratings of circuit breakers and AC and DC charging levels 1 & 2) |
| | SAEJ1773 | Standards for inductively coupled charging systems |
| | SAEJ2847 | Communication requirements between EV charging system interfaces |
| International Electro-technical Commission (IEC, Britain) [21] | IEC61851 | - Standards for EV conductive charging system operation-Cable, plug setups<br>- Onboard and offboard EMC requirements for conductive charging.<br>- Onboard and offboard charging equipment for EVs /PHEVs with 1000V AC and 1500V DC supply voltage [65]<br>- Digital communication of DC charging control between EV charging controller and supply equipment<br>- DC fast charging requirements |
| | IEC61980 | Standards for wireless power transfer for 1000V AC and 1500V DC supply voltage |
| | IEC62196 | Standards for connectors, plugs, and socket outlets used for conductive charging |
| | IEC61000-2, 3, 4 | Compatibility levels of low-frequency conductance, harmonic emission, EMC, flicker limits of voltage |
| National Electric Code (NEC) [19] | NEC625, NEC 626 | Safety measures in the off-board EV charging system (conductors, connecting plugs, inductive charging devices) |
| Underwriters' Laboratories (UL) [24] | UL2231, UL2251, UL2202 | Requirements for protection devices for EV charging circuits and charging system equipment |
| | UL2594, UL1741 | Requirements for EV supply equipment (inverter, converter, charge controller, and output controllers used in power system) |

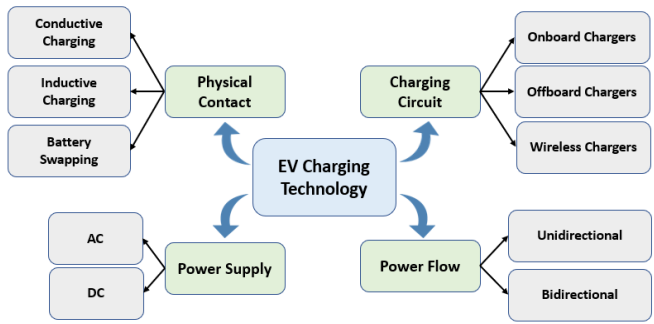

**FIGURE 4.** Classification of charging technologies used in electric Vehicles.

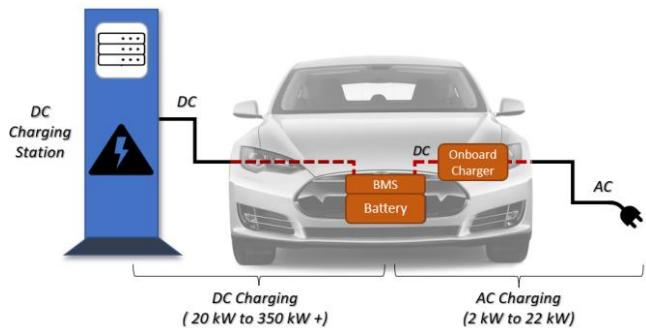

**FIGURE 5.** Onboard and Offboard charging systems of electric Vehicle

### A. GRID-TO-VEHICLE AND VEHICLE-TO-GRID MODE

The power flow direction of an EV can be either unidirectional or bidirectional according to the charging configuration built in the EV. The unidirectional charging system uses an AC-DC rectifier on the grid side and a unidirectional DC-DC converter in the onboard charger with a less complex control system. In contrast, bidirectional EV chargers can transfer power to the utility grid (discharging) as well as EV battery (charging) through offboard chargers using bidirectional AC-DC converter and bidirectional DC-DC converter [74]. The majority of the charger fleet operates in G2V mode which uses limited hardware and a simple control system to charger the battery from grid-supplied or locally generated electricity. A unidirectional charging system have simple structure which simplifies interconnection problems and tends to minimize battery degradation [75]. Unidirectional converters are executed in a single stage to reduce weight, volume, cost, and losses [76]. Moreover, active front-end unidirectional converters can offer reactive power support by controlling the phase angle of the current without discharging a battery. High penetration of unidirectional chargers can achieve power grid requirements while avoiding the cost, safety, and performance issues associated with bidirectional chargers. The comparison of unidirectional and bidirectional chargers of EV presents in table 8.

V2G mode has bidirectional energy transfer capability between EV and the electrical grid through communication strategy in charging infrastructure [77]. The bidirectional mode of EV acts as distributed generation, storage, and load for the power grid. Many researchers have recently indicated that the application of V2G in the ancillary market is more essential to voltage controlling and spinning reserve other than reducing peak load. The spinning reserve refers to the excess generation that could provide immediate backup power to the power grid. Many studies have explored EV deployment in ancillary services providing many cost-effective services and generating revenue for utilities via V2G operation. The main duties of V2G include

- Regulate battery charging operation to enhance battery life and reduce overcharging circumstances,
- Track the SoC of the battery to ensure proper charging and discharging operations and provide appropriate values of SoC and depth of discharge (DOD) to the user
- Control the EV battery SoC

The V2G can provide ancillary services including voltage and frequency regulations, improved system stability, load following, peak load shaving, energy supply, reactive power support, and RES integration. Technology improvements in EVs have introduced new energy transmission modes, vehicle to house/building (V2H, V2B), vehicle-to-load (V2L), and vehicle to vehicle (V2V). therefore, bi-directional energy transfer from EV can be categorized as below [78]

- V2G – Power flows from EV to the distribution grid
- V2H/V2B – Power flows from an EV to a home or building
- V2L - Power flows from an EV to load
- V2V– Power flows from one EV to another EV

The majority of current V2G analyses are focused on the simpler "Smart charging" control systems that extend standard demand respond applications to PEVs [79]. To thoroughly analyze V2G, recommended test programs cover three broad areas of investigation: battery impact, network operation, and system response. For V2G applications, BEVs have a high battery capacity, which results in a longer range and support for electric grid integration. For future V2G scenarios, the major areas of attention in EV development are the energy storage system, powertrain, and charging infrastructure [80].

## B. ONBOARD CHARGERS

The onboard chargers have either unidirectional or bidirectional power transfer capabilities which are compatible with level 1 and 2 chargers due to limited size, weight, volume, and power. Most of the onboard chargers use two-stage converter topologies an AC-DC stage in the front-end and a DC-DC stage in the back end [81]. Generally, a grid-connected front-end passive rectifier feeds a boost converter that operated as a PFC in onboard chargers. Then supplies appropriate power to the onboard DC-DC converter via a DC link to charge the battery [82]. The front-end rectifier stage can be achieved by a half-bridge, full-bridge, or multilevel converters. Onboard charging offers lower power transfer and therefore required more charging time compared to offboard chargers. The configuration of the onboard charger is shown in Fig. 6. Onboard charges can deliver 1.9 kW (level 1) and 19.2 kW (level 2) AC power levels. AC power is directly fed to the AC-DC rectifier in the onboard charger from the AC charging station. Then DC-DC converter regulated appropriate power levels and feeds power to the battery pack through a protection circuit by communicating with BMS and the power control unit [83].

Onboard chargers with advanced control techniques have been proposed in many research studies to improve the controllability, efficiency and grid support of the charging. In [84], a single-phase compact onboard charger with the current ripple compensator technique is proposed. The compensator consists of a zeta and boost converter which is connected in series with the EV battery in a non-isolated charging system without using bulky inductors or capacitors. A comprehensive study of wide-bandgap devices is presented in [85] for onboard chargers and demonstrated a possible approach in the onboard application via 400V/80 A test bench with Si MOSFET components. The multifunctional onboard battery charger presented in [86] can operate as AC-DC converting with PFC as well as V2G operation through sharing inductors and switches in one system. A three-phase onboard charger is integrated with the EV propulsion system in [87] by connecting the three-phase interface to the propulsion system. They system is implemented with a 3.3 kW three-phase integrated charger and proved unity power factor, 92.6% efficiency, and reduced harmonic restoration of 4.77%. Research in [88] is proposed an active power decoupling function for low-power

| FEATURES | UNIDIRECTIONAL CHARGING (G2V) | BIDIRECTIONAL CHARGING (V2G) |
|---|---|---|
| Power flow | Charging rate of EV control with a unidirectional power flow which is based on energy scheduling of G2V | G2V and V2G modes enable via bidirectional power flow to achieve a range of grid support and services |
| Type of Switches | Unidirectional power converters and diode bridge | Low and medium power transistors and high-power gate thyristor |
| Control System | A simple and easy control system, active control of charging current, and energy-pricing schemes used to manage basic control | Complex control system with additional drive circuit. Required extensive measures and an accurate communication system |
| Services | Ancillary services, load leveling, load profile management, frequency regulation [89] | Voltage and frequency regulation, backup power support during peak time, active and reactive power support, PFC and helps to integrate RESs to the gird |
| Safety | Isolated or non-isolated | Isolated or non-isolated, include high safety measures and anti-islanding protection [79] |
| Advantages | -Simple power control strategy<br>-Minimized operational cost, power losses, emission, overloading, and interconnection issues.<br>-Supply voltage and frequency regulations<br>-Provide reactive power support by controlling the phase angle of the current. | -Improve voltage profile, power quality, active and reactive power support, load leveling,<br>-Volage and frequency regulation and peak load shaving.<br>-Grid power losses and emission minimization<br>-Load factor improvement and increased profit<br>-Enable RES integration with grid |
| Limitations | -Limited services required a power connection.<br>-No extra degradation in battery | -Required 2-way power flow converters and communication.<br>-High complexity, capital cost, energy losses, and stress on the devices.<br>-Need for smart sensors and meters<br>-Fast battery degradation |

charging onboard in PEVs. The proposed onboard charger can operate in G2V, V2G, and EV battery cab be charged form the high voltage ESS by sharing capacitors, switches, and transformer in the same system.

Onboard EV charges are broadly categorized into unidirectional or bidirectional and single-phase or three-phase chargers. Various types of onboard chargers have been introduced recently as an optimum solution to the high penetration of EVs. The conventional method of EV battery charging is achieved through a dedicated onboard charger. Conventional or dedicated onboard chargers comprise two converters used for battery charging and motor controlling as shown in Fig. 7(a). Dedicated onboard chargers have limited power transfer capabilities due to several constraints including volume, cost, and weight of the vehicle [90]. Integrated onboard chargers have been designed to overcome the above limitations which are closely integrated with an electric motor using single AC-DC converter as shown in Fig. 7(b). Integrated chargers can operate the existing propulsion system for battery charging by avoiding bulky components and dedicated configurations [91]. A review of dedicated and integrated onboard charging systems is presented in the next section with commercially available onboard charger topologies.

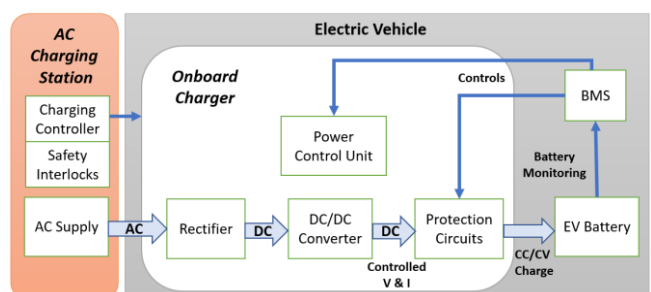

**FIGURE 6.** Configuration of conventional onboard EV charger

### 1) DEDICATED CHARGERS

The conventional or dedicated charger is an independent device with the single purpose of charging an EV battery by providing conditional output power. Dedicated is small size, lightweight and is operated with single phase or three phase AC power depending on the charging system by following level 1 and 2 charging standards. The power level has been trending upward from 3.6 kW single-phase chargers to 22 kW three-phase chargers. A dedicated charger directly connects to the AC wall socket (Mode 1 or 2) and relevant conversions such as AC-DC and DC-DC power conversions are conducted inside the onboard charger. Modern onboard chargers are following IEC 61000 standards to reduce power quality impacts on the grid. Commercial EVs have limited AC charging power levels up to 22 kW (32A and 400V three phase) due to space and weight limitations of the vehicle. The main challenges of onboard chargers are dependence on the charging outlet, voltage limitations of battery, DC current controlled with the AC voltage controller, and incompatibility of ground referenced. Moreover, extensive safety requirements need to be addressed at high power levels and the size and weight of the vehicle may increase as increasing addition components. Dedicated DC chargers (22 kW) are installed in houses, workplaces, apartments, and shopping centers.

Most of the commercially available onboard chargers have two-stage power converter topologies. The usual onboard charging configuration includes an electromagnetic interference filter, AC-DC converter, and isolated DC-DC converter. Grid side AC-DC converters are comprised of a PFC circuit to limit harmonics and supply power to the DC link as a first stage and then the DC-DC converter is connected to the battery interface which is generally comprised of two inductors and a capacitor (LLC) with two or four switches to supply highly efficient power transfer. A large capacitor is required in between two converters to filter grid frequency. Various onboard charger topologies and control systems have been reviewed in [92], [93], [82]. The onboard converter topology of the 2016 Volt is shown in [94]. Most conventional configurations of onboard chargers are interleaved PFC boost converter and diode bridge converter. According to General Motors' evaluations, interleaved topologies are widely used in front-end conversion stages for modern onboard chargers [95]. The second generation Volt onboard charger is shown in Fig. 8(a) which comprises four diode bridges and two parallel interleaved boost converters on the grid side to enable approximately 400 V of intermediate DC link voltage. Resonant LLC full bridge converter is used for the DC-DC conversion stage to acquire output voltage for the battery.

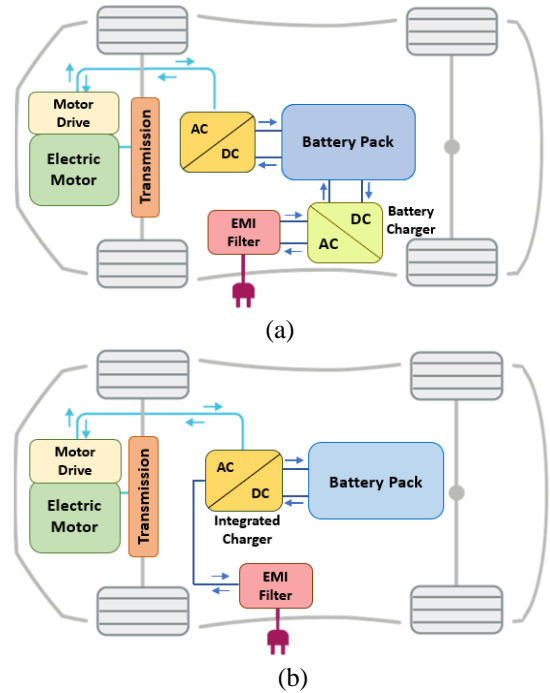

(a)

(b)

**FIGURE 7.** Configuration of onboard power electronic interface (a) Dedicated onboard charger (b) Integrated onboard charger

The new version of the Tesla onboard charger adopted a similar trend in the DC-DC conversion stage and three parallel channels are integrated into the front end as shown in Fig. 8(b). The maximum charging capability of Tesla onboard chargers is 11.5 kW and 240kW V3 superchargers are used for EV battery charging in Tesla [96], [97]. Fig.8 (c) shows the topology of Hyundai onboard chargers which can support vehicle-to-device applications via bidirectional power flow. The onboard charger is comprised of a front-end single-phase active rectifier, and a bidirectional buck-boost converter followed by a ducal active bridge to facilitate bidirectional power flow as well as adjust the appropriate voltage for different configurations [98]. Bridgeless boost type PFC topologies are also used in dedicated onboard chargers by replacing passive diode rectifiers to reduce conversion stages and power losses [99]. Matrix-type converters have been introduced by Hella electronics by further reducing the conversion stages. The matrix converter turns the input grid frequency into the intermediate frequency and a large DC filter connects to the battery side in the Hella electronics onboard charger shown in Fig. 8(d). The maximum efficiency can be achieved up to 98% for a 7.2 kW single-phase operation in this converter [100].

## 2) INTEGRATED CHARGERS

Integrated chargers have been designed to overcome the limitations of conventional onboard chargers while preserving their advantages including fast charging capability, reduced components, cost, and volume of the charger [101]. Integrated onboard chargers utilize a propulsion system, electric motor, and inverter for EV battery charging by avoiding separate convention stages with bulky add-on capacitors and inductors. Therefore, they can offer bidirectional high-power levels (Level 1 and 3) and more space for the battery [102]. The propulsion inverter operates as a bidirectional AC-DC converter and motor winding provides galvanic isolation and filter conductance [103], [104]. Split-winding AC motors are used in non-isolated integrated chargers. However, single-stage integrated chargers may have current ripples at the DC side and need addition components to reduce voltage ripples. The traction controllers may limit the charging power and the electric motor maybe operate in charging mode in integrated chargers and technical requirements such as motor winding limitations, and zero average torque may exist [105]. Renault pioneered integrated charging design and Ford Motor company currently uses an integrated onboard charger that combined battery charging and motor drive based on an induction motor. Renault is pioneered in integrated chargers [106],[87].

Most integrated chargers inversely use the electric drive inverter as a boost stage with more than 50 kW power, and it can utilize the propulsive components in the charging period. Different types of integrated charging topologies have been proposed in recent years using general DC-DC converter, switched reluctance motor, or alternating motor [63]. In [75],

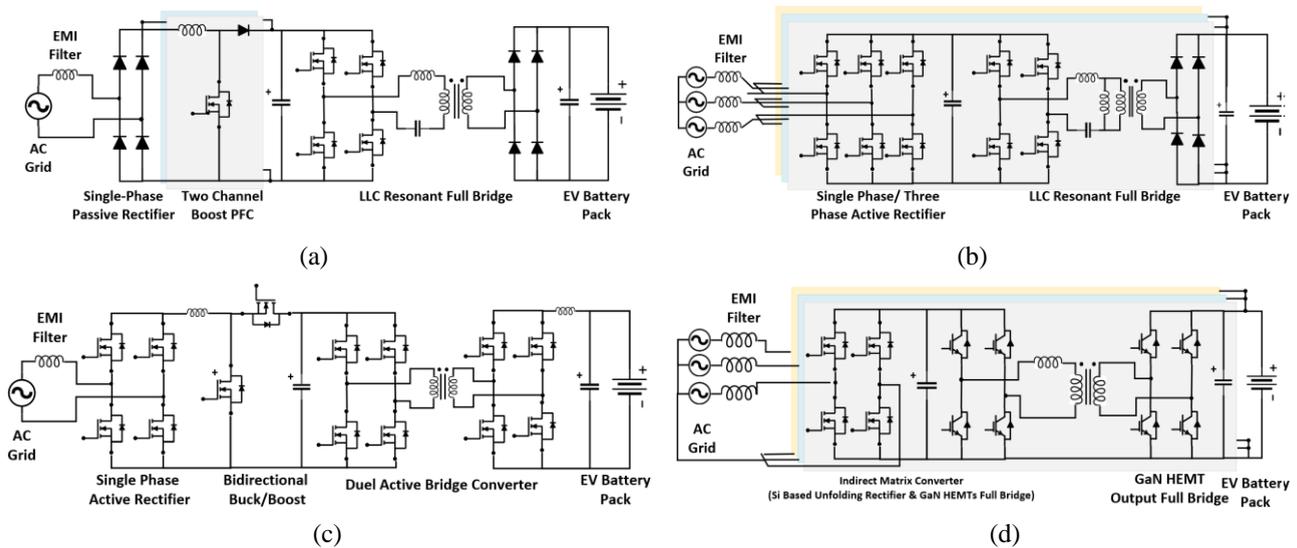

(a)

(b)

(c)

(d)

**FIGURE 8.** Configurations of dedicated onboard charger (a) Second-generation Volt, (b) Tesla Model 3/Y, (c) Hyundai vehicle-to-device, (d) Hella Electronics/GaN Systems

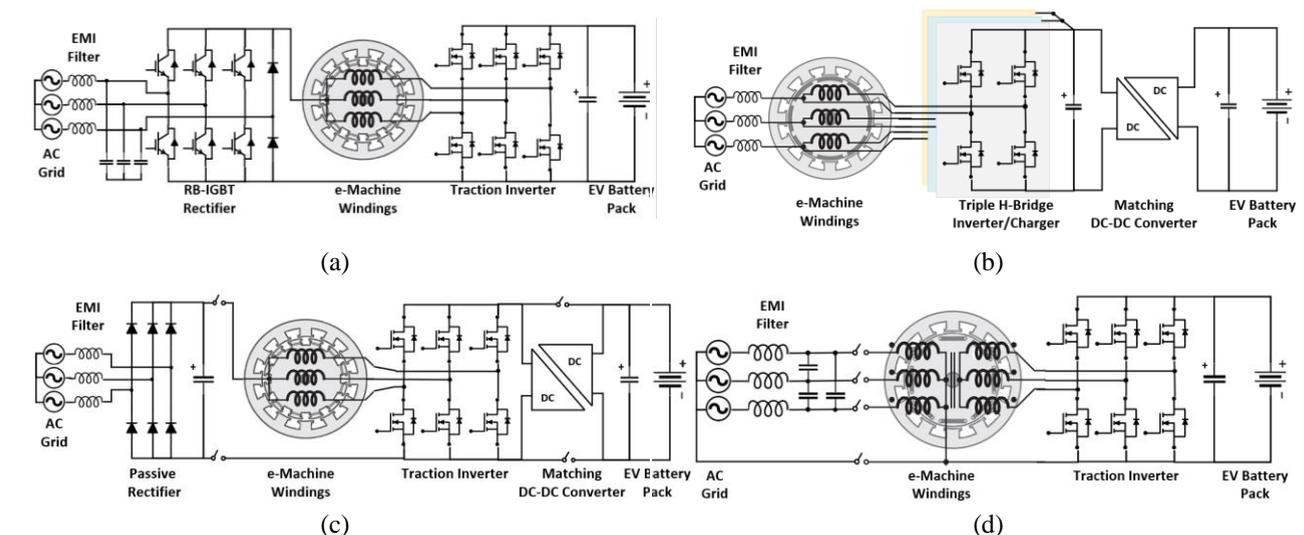

(a)

(b)

(c)

(d)

**FIGURE 9.** Configurations of Integrated Onboard Charger (a) Renault Chameleon, (b) Valeo dual-inverter charger, (c) Continental All Charge System, (d) Galvanically isolated traction integrated charger [11], [91]

TABLE IX
SPECIFICATIONS OF COMMERCIALLY AVAILABLE ONBOARD CHARGERS [114], [115]

| MODEL AND MANUFACTURER | BATTERY CAPACITY (kWh) | CHARGING POWER (kW) | BATTERY VOLTAGE (V) | CHARGING TIME (Minutes) | DRIVE RANGE (kM) |
|---|---|---|---|---|---|
| Model S, long range  - Tesla - 2022 | 100 | 200 | 400 | 24 | 624 |
| Model 3 Performance  - Tesla - 2021 | 79.5 | 120 | 360 | 33 | 567 |
| Bolt EUV  - Chevrolet - 2022 | 65 | 50 | 350 | 66 | 402 |
| Leaf SL  - Nissan - 2019 | 62 | 100 | 360 | 35 | 346 |
| Leaf S  - Nissan - 2019 | 40 | 50 | 350 | 36 | 378 |
| Ioniq  5 Long Range  - Hyundai - 2022 | 72.6 | 160 | 800 | 18 | 412 |
| e-208 GT  - Peugeot - 2019 | 50 | 100 | 400 | 30 | 450 |
| Taycan 4S  - Porsche - 2022 | 79.2 | 225 | 800 | 21 | 407 |
| MX-30  - Mazda - 2021 | 35.5 | 50 | 355 | 34 | 265 |
| e-tron 55 Quattro  - Audi - 2022 | 95 | 150 | 396 | 26 | 441 |
| Q4 Sportback 55  - Audi – 2022 | 82 | 110 | 400 | 38 | 460 |
| i4 M50  - BMW – 2022 | 83.9 | 210 | 398.5 | 31 | 510 |
| iX xDrive50  - BMW -2022 | 111.5 | 195 | 330 | 35 | 630 |
| EQS 350  - Mercedes Benz - 2022 | 90.56 | 170 | 500 | 30 | 626 |
| I-Pace S AWD  - Jaguar - 2020 | 90 | 100 | 388 | 43 | 470 |

integrated converters are comprehensive analyses based on the motor type either isolated or non-isolated cases. The topology of the Renault Chameleon integrated charger is shown in Fig. 9(a) which employs a reverse blocking IGBT rectifier with filtering components at the AC side for single and three-phase AC grid [107]. As the first commercially used first integrated charger, Chameleon chargers are currently used in Renault ZOE which is not required addition components between operating modes as torque is not generated in the motor [108]. The motor winding is act as a DC link and the traction inverter is connected between the motor and the battery to supply the required current for the battery pack [109],[110]. Specifically, the Chameleon charger is designed to use the neutral point of the motor to turn the motor inverter into three separate boost-type converters [82].

Configuration of motor winding or additional grid to motor interface helps to enable high power charging without producing torque in the electric motor [82]. Valeo integrated charger is developed using a triple H-bridge inverter which is connected using a triple H-bridge inverter which is connected to spilled winding in the synchronous motor as shown in Fig. 9(b). The inverter can provide high voltage at an intermediate DC-link and a matching DC-DC converter is implemented between the battery and inverters to adjust the battery voltage. Passive rectifier and filter components are additionally used in Continental high-power onboard chargers as shown in Fig. 9(c). A high power charging rate is possible with a three-phase current (400 V) and galvanic isolation provides additional protection in Continental onboard chargers [111]. The multiphase integrated onboards charger presented [82] has decoupled inductors in the motor and a multiphase AC-DC converter that is directly connected to the battery. overview of multiphase integrated onboards chargers can be found in [112], [103]. Another type of integrated charger is an isolated onboard charger which is equipped with multiterminal motors to execute traction mode as well as galvanic isolation during battery charging. Isolated integrated charger topology is shown in Fig. 9(d) which included two sets of three-phase motor winding. Stator windings are normally connected in series to form a three-phase set depending on the charging configuration. This dedicated chargers can used for single-phase systems as well as operates as a high power isolated bidirectional fast charge is unity power factor [113]. Commercially available onboard chargers are presented in table 9 with specifications including battery voltage, capacity, charging power, charging time and driving range. Most modern electric cars utilized with high voltage battery (up to 800V ) and hence isolated high-voltage transmission system is added for safety.

## C. OFFBOARD CHARGERS

Offboard chargers are integrated with DC fast charging or ultra-fast charging systems for high power flow (>20 kW) between the utility grid and battery based on level 3 or extreme fast charging standards. Power conversion stage of the offboard charger is located outside of the EV and therefore volume, weight, size, and cost of the charger are significantly reduced when compared to the onboard charger [116]. As a conductive charging process, EV offboard chargers are incorporated with either AC bus or DC bus configuration [34]. Most fast and ultrafast charging systems prefer AC bus-connected fast charging stations due to well-equipped configurations and matured power converters on the AC power grid. They consist of two converter stages AC-DC and DC-DC converter to adjust DC current before eaching the EV as shown in Fig. 10. Central AC-DC converter is connected to the low-frequency transformer on the grid side in the DC bus connected offboard charging systems. The DC-DC converter is connected to the DC link to provide DC power to the battery. DC bus-connected systems are more efficient and flexible than AC bus-based fast chargers and RESs can be connected via DC link and grid side impacts are simply avoided [34]. Moreover, AC and DC bus-based configurations are also available for offboard chargers. However, fast charging stations have some drawbacks including high infrastructure costs, safety requirements, complex control strategies, and communication protocols that need to be used according to the standards [117]. Review of offboard charger topologies and control techniques have been presented in [75], [79], [118], [119].

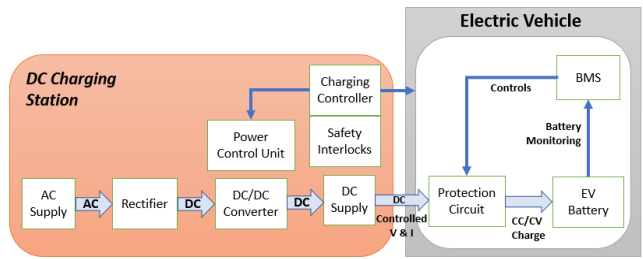

**FIGURE 10.** Configuration of Conventional Offboard Charger

Modern offboard chargers can provide more than 350 kW power to the EV battery for ultrafast charging and are compatible with 800 V EVs in near future. Most of the offboard charging topologies are employed with galvanic isolation in the DC-DC converter stage using a high-frequency transformer (50 kHz – 300 kHz) instead of a line-frequency transformer to provide safety for the components, better control of voltage adjustments, and compactness [75]. Most EV manufacturers tend to design chargers by modularizing to achieve compatibility, high efficiency, and economic benefits from their chargers. Terra 53/54 series fast charger is designed based on a power electronic building block (PEBB) by ABB as a benchmark for offboard chargers as shown in Fig. 11(a). The PEBB is a widely accepted concept which incorporate number of topologies to reduce cost, size, losses, and components of the applications [121]. The number of active power stages varies with output requirements and an isolated DC-DC converter is used to meet high power level and isolation requirements. The modular system of ABB Terra 53/54 fast charger is designed by replicating the same type of PEBB (5 × 3 PEBBs to Reach 50 kW) and efficiency is 94% [122],[123]. ABB Terra high power charger series is shown in Fig. 11(b) which is also configured with the modular system (three PEBB to reach 150 kW high power). An isolation transformer and LCL filter are used to reduce grid side harmonics and an active rectifier and interleaved buck converter are used in the modular

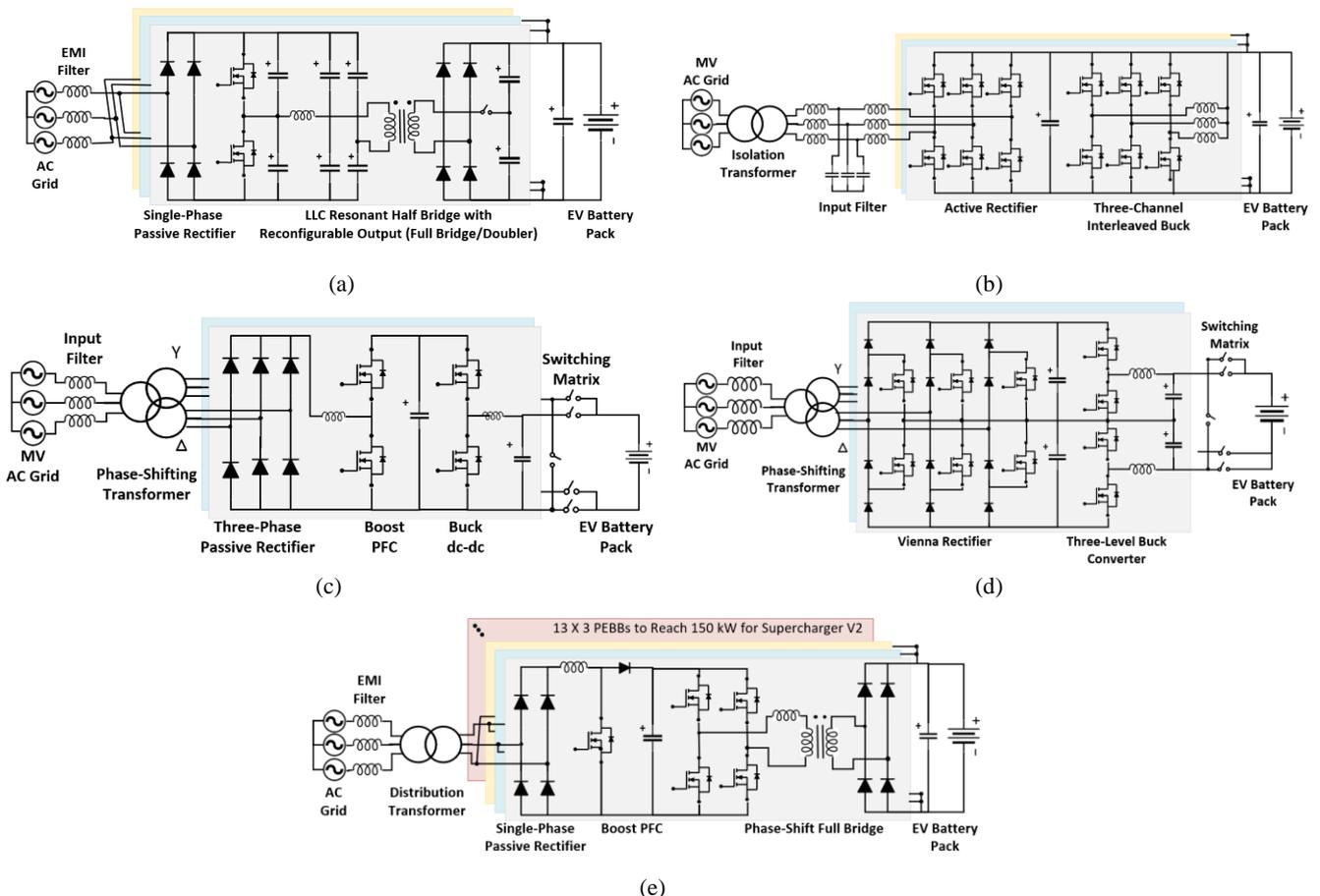

**FIGURE 11.** Configurations of Offboard Chargers (a) ABB Terra 53/54 50-kW fast charger, (b) ABB Terra HP 150-kW high-power charger (c) Porsche Modular Fast Charger Park A (d) Porsche Modular Fast Charging Park B (e) Tesla V2 Supercharger [120],[11]

configuration in high-power ABB Terra offboard charger.

Porsche fast charger can manage 800V with modular fast charging topology. Porsche Modular Park A and park B fast charging configurations are shown in Fig. 11(c) and (d) respectively. A phase shifting transformer is used after the input filter to provide galvanic isolation and improve the power quality of the AC grid side in both configurations. Three-phase passive rectifier and boost PFC converter used after phase shift transformer and DC-DC buck converter utilized to lower current ripple and stepdown voltage as required for EV battery in Porsche Modular Park A fast charger [124]. A combination of Vienna converter and three-level interleaved buck converter enable modifications in battery charging converter to provide PFC, reduce current ripples, and be compatible with other configurations in Park B fast charger [125], [126]. DC fast chargers are still in the developing phase and therefore standards and protection requirements are not well established due to specifications in high power, complex grounding topologies, and fault types [127]. Moreover, protection and metering requirements are critical for bidirectional fast chargers as they are very sensitive to grid disturbances and a review of coordination techniques is presented [128]. Tesla superchargers have a combination of PEBBs (13 x 3 PEBBs) to provide 150 kW power with 92% efficiency as shown in Fig. 11(e). The simplified one-line diagram of the Tesla supercharging station is presented in [122]. Specifications of currently available ultra-fast and fast charging systems are presented in Table 10.

## VI. CONVERTER CONFIGURATIONS OF EV CHARGING SYSTEMS

The power electronic converters are an integral part of electrification to achieve efficient, and reliable operation of EV charging systems. As advances in power electronic techniques have made conversions possible to achieve cost-effective and maximum power conversion. The power converter topologies are interfacing between the EV battery system and power network which is responsible for improving charging performances and controllability of the charging system [129]. Moreover, advanced power converters and controller are continuously developing with the increasing integration of EV charging systems into the RESs in recent years. The AC-DC converters and DC-DC converters are equipped with EV charging systems to supply power to the vehicle components and deliver power from a power grid to an on- board high- voltage battery pack. The AC-DC converters are mainly used for rectification and power factor correction (PFC) of the EV charging system and are further able to control charging cost, and complexity, and improve power quality. The front end of the converter consists of the AC-DC converter to convert single/three-phase power to DC power and supply the required DC link power and function as a power factor corrector topology. The DC-DC converters are primarily employed in an EV charging station, classified as unidirectional or directional converters and isolated or non-isolated power converters [93].

The significant advantages of EVs over many other conventional clean energy applications are largely due to improvements in converter topologies. The improved power converters in EV charging systems can operate bidirectionally concerning the power demand of the grid, improve power quality and provide ancillary services. The simplified diagram of a conventional EV system is shown in Fig. 12 which comprises with DC-DC converter, AC-DC converter, and filter in between the EV battery and grid. The AC-DC converter rectifies AC voltage to regulated intermediate DC link voltage and the DC-DC converter controls DC input voltage for the EV battery. The control system is maintained the transient and steady-state performance of the system by providing relevant control signals for converters. Moreover, PFC techniques are implemented parallelly with the AC-DC converter to achieve a unity power factor and overcome current harmonic impacts [130]. The PFC circuit is sensing input voltage and current and then controls the input currents close to sinusoidal and in phase with relevant voltages by controlling the converter switches. The desired DC link voltage is regulated via a DC-DC converter to charge the EV battery.

### A. AC-DC CONVERTERS

The AC-DC converter provides the interface between the DC link and the power grid by providing high power quality on the DC and AC sides of the charging system. They are generally designed as single-phase H-bridge inverters or three-phase three-leg inverters either controlled or uncontrolled rectifiers [71]. The AC-DC conversion is the first stage of the EV charging system, and it controls reactive power consumption and grid-side current harmonics [131]. The PFC technique is implemented in the AC-DC converter to supply efficient and safe power output to protect the connected devices, users, and the grid. The AC-DC converters can be classified as unidirectional and bidirectional as well as single-stage and multistage AC-DC power converters. The unidirectional chargers deliver G2V operation (charging) with a low-cost and simple charging structure and supply moderate grid support with minimum power infrastructure modernizes [132]. Conversely, bidirectional converters can offer G2V and V2G (charging and discharging) operation with advanced and coordinated control between EV, charging station, and grid with a high level of ancillary services.

**FIGURE 12.** Block Diagram of Conventional EV charging System

TABLE X
SPECIFICATIONS OF CURRENTLY AVAILABLE ULTRA-FAST AND FAST AND CHARGERS

| MODEL | INPUT VOLTAGE (Vac) | OUTPUT VOLTAGE (Vdc) | OUTPUT CURRENT (A) | POWER (kW) | SUPPORTED PROTOCOLS |
|---|---|---|---|---|---|
| ABB Terra 54 [133] | 400 Vac +/- 10 % Three-Phase | 150-500 | 125 | 50 | CHAdeMO CCS |
| ABB Terra High Power GEN III [134] | 400 Vac +/- 10 % Three-Phase | 150-920 | 500 | 350 | CHAdeMO CCS1, CSS2 |
| Tesla Supercharger V3 [135] | 380 - 480 Three-Phase | 880-970 | 640 | 250 | Superchargers |
| Signet FC100K-CC [136] | 480 Vac | 150-500 | 200 | 100 | CHAdeMO SAE Combo |
| Tritium PK350 [137] | 480 Vac Three-Phase | 200-920 | 200-500 | 350 | CHAdeMO CSS2 |
| Blink 60kW DCFC [138] | 480 Three-Phase | 150-500 | 140 | 60 | CCS1 |
| Blink 180kW High Power DCFC [139] | 480 Three-Phase | 150-1000 | 240 | 180 | CCS1 |
| EVBox Troniq 100 [140] | 400 Three-Phase | 50-500 | 200 | 100 | CHAdeMO CCS2 |
| Siemens VersiCharge Ultra 175 [141] | 480 Vac +/- 10 % Three-Phase | 200-920 | 200-350 | 178 | CHAdeMO CCS |
| Ingeteam - INGEREV RAOID ST400 [142] | 380-440 Vac Three-Phase | 50-920 | 500 | 360 | CHAdeMO CCS |

The single-stage AC-DC converter is combined with a DC-DC converter which reduced expensive components including DC link capacitors and inductors in the system. However, a single-stage converter has less output voltage range and non-isolated converters have a limited conversion ratio [92]. The single-stage modular three-phase AC-DC converter is introduced in [143] to voltage regulation and PFC and single phase isolated AC-DC converter is proposed in [144] from differential boost converter by using the AC decoupling waveform technique to address reliability issues. In contrast, multistage converter topologies are designed with two or more converter levels and high-power levels of the converter are provided efficient and reliable control for the charging system. The two-level and three-level voltage source converters are widely used for charging applications including buck/boost converters and multilevel converters. Moreover, the filter is connected between the AC-DC converter and power grid to reduce harmonics, di/dt on semiconductors, and isolate the converter from the power grid [145]. The commonly used filters are LC and LCL filters for AC-DC converters and more advanced filters are used for fast and ultra-fact charging stations.

### 1) BUCK-TYPE RECTIFIER

The buck type of converter is used to regulate the output voltage which is lower than the input voltage with unidirectional power flow. The three-phase buck converter has a wide range of features in the AC-DC power stage when compared to the three-phase boost type converters. They have a wider voltage control range, inherent inrush free direct startup, allow dynamic current limitation at the output, provide overcurrent protection during short circuits, and can maintain PFC capability at the input side [146],[147]. The buck converter-based PFC topologies can be classified as bridgeless, interleaved, and bridgeless interleaved buck converters [148]. The conventional six-switch three-phase buck converter shown in Fig.13 (a) includes three legs that are connected to the three phases and one freewheeling diode to lower the conduction losses during the freewheeling condition. The freewheeling diode is divided into a series connected by two diodes and the common node is connected to the input neutral point in the study [149] to reduce voltage stress on the converter switchers. Conventional buck converters are employed for low-power charging systems (< 300 W) due to their capability to provide improved power quality and efficient at different line voltages [148]. The input filter is critical for the buck converter as shown in Fig.13 (a) to reduce the inherent input current disturbances (high ripple) from the charging system [150].

The bidirectional five-level buck converter has been proposed in [151] which is employed two voltage sensors with a complex control strategy to balance voltages across the two capacitors and high voltage side power switches ratings are equal to twice the DC voltage output. The distributed parasitic capacitance of the high-density three-phase buck converters is a major challenge in high-frequency operation which lead to input current distortion and an increase in THD mainly under light load condition. The modified three-phase buck converter has been presented in [152] to reduce the impact of distributed parasitic capacitance between the DC link output and the system ground.

Moreover, high stepdown voltage gain may appear when the multiple EVs charging due to variation in the range of the EV battery. Furthermore, power quality impacts and losses increase when the voltage output is less than three quarter of input voltage due to decreasing modulation index less than 0.5 of the standard buck converter. The matrix-based theephase buck converter has been implemented in [153],[154] to regulate the modulation index and improve grid support.

### 2) BOOST-TYPE RECTIFIER

DC voltage output, low current stress and THD, bidirectional power flow, and high efficiency with a simplified control scheme. The boost converters are integrated with PFC configuration for EV charging systems which are generally operating in continuous conduction mode which is selected for medium and high-power applications. The boost converter exhibits lower conducted electromagnetic interference (EMI) than other buck and buck-boost converters because of continuous current flow capability [155]. The main limitation of this converter is high conduction losses due to the current flowing via semiconductor components and the diode recovery losses are imposed by the high-frequency operation of the converter [116]. The three-phase six-switch boost converter shown in Fig. 13(b) consists of six switches in the there-legs and an LC filter to reduce input current harmonics and boost the voltage. The switchers upper and lower are executed in complementary and inductors are employed to boost the voltage and reduce input current harmonics. The three-phase three-level boost converters can balance the input AC system during unbalanced input voltage and reduce harmonics at the DC link voltage by employing a bulky capacitor or developing an active control method [156]. The power losses increased in conventional boost converter topology at a high power rate due to high ripple occurs at the output capacitor.

The EV charging systems incorporate a variety of boost converter topologies, including bridgeless, interleaved, and bridgeless interleaved boost topologies. In addition, asymmetrical, and symmetrical bridgeless boost rectifiers have enhanced efficiency when compared to the regular boost converter due to the less operating electronic devices. The semi-bridgeless boost rectifier is presented in [157] for font-end AC-DC converter of PHEV charger to minimize the charger size, cost, and EMI and increase efficiency at light load. The isolation approaches such as power supplier separation or transformers are generally used in high power applications to avoid current circulating which may increase the volume, passive components, and cost of the system. The parallel three-phase boost converter circuit has presented in [158] with the potential of zero-sequence current circulating capability, modular design, and high efficiency. The harmonics of the unbalanced ac input voltage can be mitigated by adding a bulky capacitor and improving active control techniques to reduce harmonics in the DC-link voltage [159]. The magnetic circuit effects and size of the converter can be reduced by an interleaving of boost-type two converters which doubles the switching frequency and improved energy efficiency [76].

### 3) SWISS RECTIFIER

The Swiss converter is a buck-type PFC converter topology suitable for EV charging systems with 250 - 450V DC bus voltage and 380 V three-phase AC voltage [160]. The Swiss converter has low common mode noise, switching losses lower complex power circuits, control strategy, and inherent free inrush limitation [161],[162]. The Swiss converter is generally implemented with three phase unfolder circuit and two DC-DC buck converters. The Swiss converter's three-phase unfolder electric circuit uses two full bridge circuits to transform the AC voltage into time-varying two positive voltages. As a result, less high-frequency transistors are required than in single-stage isolated converters. The schematic of eight switches Swiss converter is shown in Fig. 13(c) comprises a uncontrolled three-phase converter bridge and three sets of low-frequency bidirectional switchers Sya, Syb, and Syc which can be defined as six volage segments concerning the frequency of the phase voltage. The two active switches T+ and T- operate corresponding to the two-phase voltage which is involved in generating output voltage [34].

The single-stage full-bridge Swiss converter is presented in [160] and the midpoint clamper is used to integrate the PFC method of the converter. The system achieved 95.4% efficiency under half-rated power in a 10 kW system and showed 5% input current THD under-rated power. The higher switching frequency or increased AC input filter are used to decrease voltage and current ripples at the input. But those options may increase the volume, cost, and losses of the system. The interleaving of Swiss rectifiers can be used to overcome the above drawbacks and offer high reliability, power, and bandwidth, low current and voltage ripple at the input and output, and lowering filter requirement. The three-phase Swiss converter with interleaved DC-DC output has been presented in [163] and the efficiency of the system is 99.3% in the 8 kW rated power. The multilevel Swiss rectifiers are also used for high power applications but the control scheme becomes complex [164]. Moreover, bidirectional Swiss converters can be incorporated with the smart coordinate controller to operate V2G in EV charging systems [165].

### 4) VIENNA RECTIFIER

The Vienna converter is used to supply controlled DC bus voltage in high power applications and performs as a three-phase boost-type PFC rectifier. Vienna converter provides many advantages when compared to the other three level converters such as high-power density, efficiency, stable voltage output, reduced number of switches, low voltage stress of the semiconductor, lower THD, unity PF, and neutral connection free structure [166],[167]. Conventional control methods of Vienna converter are sliding mode variable strategy, hysteresis current control, and double closes-loop control techniques which can be used to regulate voltage of the DC-link and unity PF. Moreover, the dead zone is not required to drive switches and voltage stress on the switches is appeared on half of the two-level converter at the same DC link voltage [168]. The schematic diagram of three phase Vienna converter is shown in Fig. 13(d). The converter is consisting of three inductors for boost state at the input side, three power bridges for three phases, and two series output split capacitors on the DC link. Power flow of this converter is unidirectional and each power bridge comprises two fast rectifier diodes and two reverse series connected switchers.

The Vienna converter is designed to enhance the large-scale integration of EVs on the grid [169] using a virtual synchronous machine control strategy. The sliding mode control loop method is utilized in a three-phase AC-DC Vienna rectifier in [170], which consists of loss-free resistor behavior in each phase for PFC. The three-phase interleaved Vienna rectifier is implemented in [171] by focusing on switching frequency circulating current generation with interleaving control. The efficiency of the converter is 99.98% in a 3 kW prototype at normal load conditions. Furthermore, a comparison of power losses of different Vienna converter-based configurations is analyzed in [171] and it was found that the lowest power loss belongs to three phase Vienna converter. In [34] bidirectional Vienna converter is implemented for V2G operation by replacing six fast rectifier diodes of Fig. 13 (d) with switchers to modify it as a T-type PFC configuration. The bidirectional T-type Vienna converter has higher efficiency, and lower conduction losses and it is suitable for V2G operation and storage applications. Additionally, the Vienna converter works with bipolar DC-bus structures, which improve power flow capabilities while lowering DC-DC power stage step-down ratio [172].

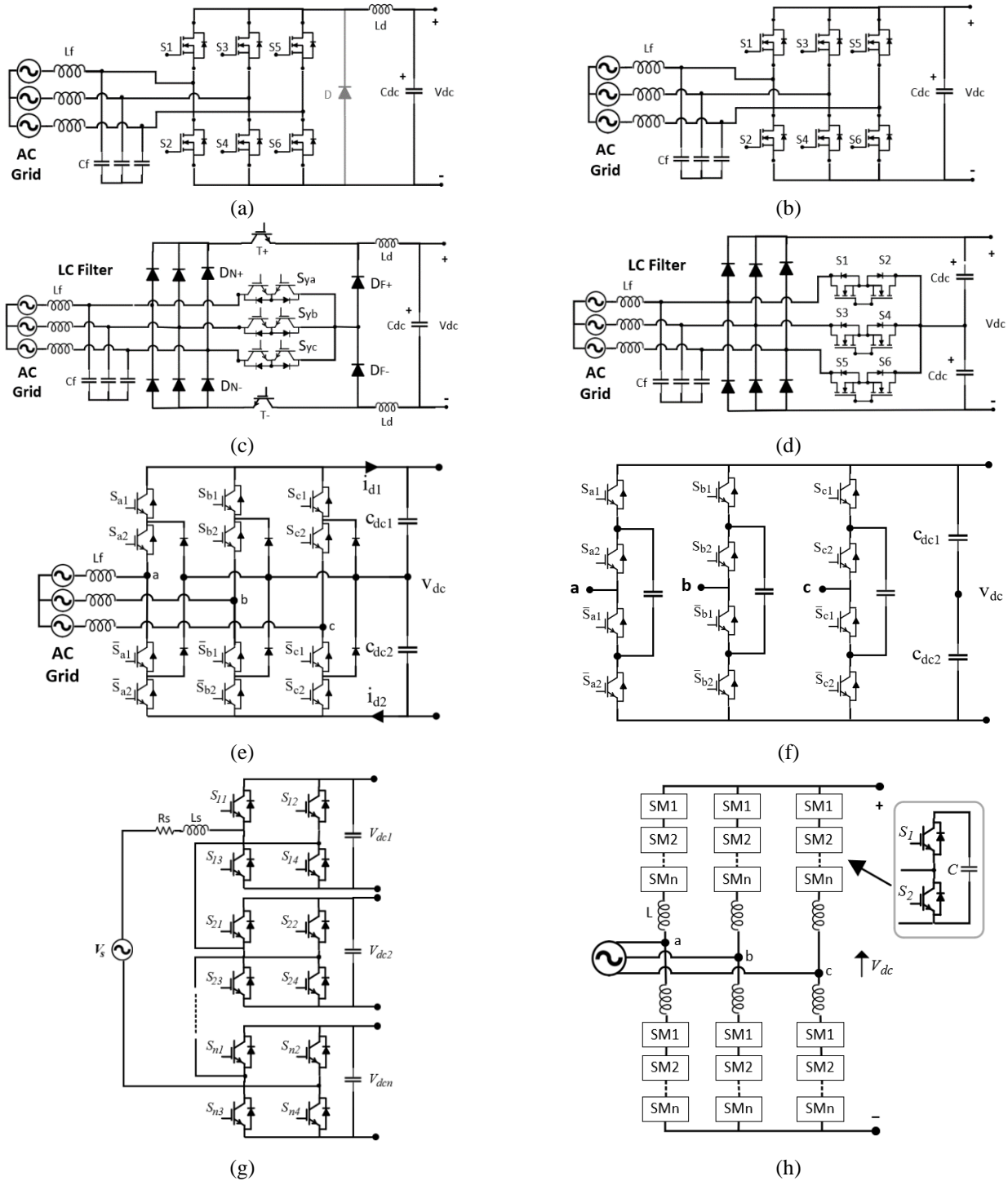

**FIGURE 13.** AC-DC Converter Configurations (a) Three-phase six-switch buck-type converter, (b) Three-phase six-switch boost converter, (c) Three-phase Swiss Converter, (d) Vienna Converter (e) Three-level neutral point clamped (NPC) converter, (f) Three-phase three-level flying capacitor inverter (g) Cascaded H-Bridge (CHB) multilevel active rectifier (h) Modular Multilevel inverter [76], [34].

## 5) MULTILEVEL AC-DC CONVERTER

Multilevel converters are widely accepted for AC-DC conversion in fast and ultra-fast charging applications over other converters due to several reasons. The multilevel converter concept is developed for high voltage and power applications with the ability to unidirectional and bidirectional power flow, transformer-less operation, and high quality outputs [173],[174]. The level 3 EV charging systems comprise multilevel converters as they provide high efficiency and power density as well as supply alternating voltages from various lower dc voltages [175]. The functionality of the multilevel inverter is depending on either an isolated DC source or a series of connected split capacitors which are connected to the single DC source to provide sub-level voltage outputs [176]. The multilevel converter topologies can be categorized as cascaded converters, neutral point clamped (NPC), and flying capacitors as shown in Fig. 14. The cascaded multilevel converters can be divided into cascaded H-bridge (CHB) and modular multilevel converter (MMC). The hybrid multilevel converters are designed by using two or more of the mentioned converter topologies [177].

The neutral point clamped converter is used in low and medium voltage operation applications which can reduce harmonics, dv/dt stress across converter switches, enhance the power capability of the EV charging stations, and reduce step-down effort by DC-DC charger [178]. The three-phase 3-level NPC multilevel AC-DC converters are shown in Fig. 13(e) and the switching loss can be reduced by blocking all switches to half of the DC-link voltage. The NPC with a central AC-DC converter is designed for the EV charging system in [179] via medium voltage grid and bipolar DC bus. However, uncertainties of random EV connections and unbalanced problems in bipolar DC buses are unable to control with the NPC modulation stage. Therefore, an additional circuit was added to the NPC converters to voltage balancing with the three legs in [180]. The next multilevel type is a flying capacity multilevel converter which required lower component volume, used low voltage switches, and has has fewer losses than other converters [181]. The schematic of three phases three-level flying capacitor multilevel converter is shown in Fig. 13(f). The flying capacitor with closed-loop control techniques is presented in [182] and six levels interleaved flying capacitor converter is proposed in [183] to achieve high power density and efficiency.

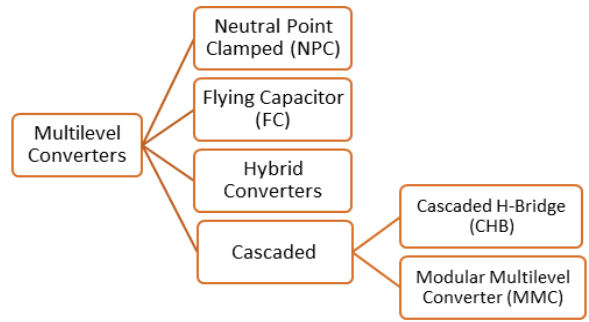

**FIGURE 14.** Classification of Multilevel Converters

The cascaded H-bridge multilevel converter is comprised of a series of connected H-bridge (full bridge) cells that are coupled in cascade on the AC voltage side [184]. The output waveform of the synthesized multilevel converter includes more steps as the level count increases, which generates a staircase wave with the intended waveform [185]. The modular structure of the multilevel converters or modular multilevel converters (MMC) is the most attractive AC-DC converter used in EV and RES applications. The cascaded H-bridge multilevel and modular multilevel converters are shown in Fig. 13(g) and (h) respectively. The MMC is playing a major role in the industry due to its advantages over other converters [186]. The MMC is commonly used in RES and EV charging systems to mitigate large-scale grid integration and improved power quality. The large EV or HEV drives and EV charging stations are utilized with modular multilevel AC-DC converters [175]. A novel MMC topology was proposed in [187] which compensates for voltage imbalances.

## B. DC-DC CONVERTERS

The DC-DC converter is directly connected to the EV battery and required specific features to integrate with the high voltage DC link of the charging systems. It converts intermediate DC voltage into a desired regulated DC voltage level to charge the EV battery pack. The DC-DC converters can be divided into unidirectional and bidirectional or isolated and non-isolated according to the requirement. They are light-weighted, highly efficient converters and have high voltage gain, a wide range of power delivering power, and less passive components [188],[189]. The isolation is essential for high power fast charging stations which are generally granted in between the EV battery and the grid such as a line-frequency transformer connected in front of the AC-DC converter or a high-frequency transformer connected to the DC-DC converter. However, high voltage DC-DC converters have some disadvantages such as low efficiency due to hard switching, difficult in designing high bandwidth control loops, inability to achieve high power density, and providing flat voltage with moderate switching gate signals [190], [188], [191].

The DC-DC conversion is the second stage of EV charging systems which involved additional requirements on the battery side such as charging speed, power controlling, isolation, or non-isolation. Moreover, the converter is performing as a PFC circuit to minimize THD, and current harmonics and maintain unity PF. The common non-isolated DC-DC converter topologies include conventional, interleaved, and multilevel converters, and isolated DC-DC converters comprised of half-bridge, full-bridge, Z-source, forward, flyback, and multiport converter [192], [193]. The current and voltage-fed full-bridge converters are the extensively used configuration for EV charging systems. The ZVS (Zero voltage switching) circuit is gained on the side of the voltage-fed converter and ZCS (Zero current switching) circuit is acquired on the side of voltage-fed converters. In addition, the dual active full bridge with voltage-fed converters is widely used for the primary and secondary sides of the converter [34].

### 1) ISOLATED DC-DC CONVERTERS

The dual active bridge (DAB) converter is commonly used for EV charging systems due to its high efficiency, reliability, bidirectional power flow, buck and boost capability, low stress on components, and small filter requirements [194],[195]. The DAB converters are preferred for many applications as they have improved power density and converter efficiency based on the new SiC and GaN-based semiconductor devices [196]. The converter has a fixed switching frequency and small passive components, as well as primary and secondary bridges of the converter, which is generally controlled by ZVS operating due to simplicity. The schematic diagram of the DAB DC-DC converter is shown in Fig. 15(a). The DAB DC-DC stage is utilized to achieve sinusoidal waveform by adjusting the phase shift between primary and secondary voltage. The bidirectional power flow can be simply achieved from DAB converters when compared to the other resonant converters. The power flows from left to right when the positive phase shift angle is measured in between input and output bridge voltage and enables the reverse power flow by changing the phase shift angle polarity [197].

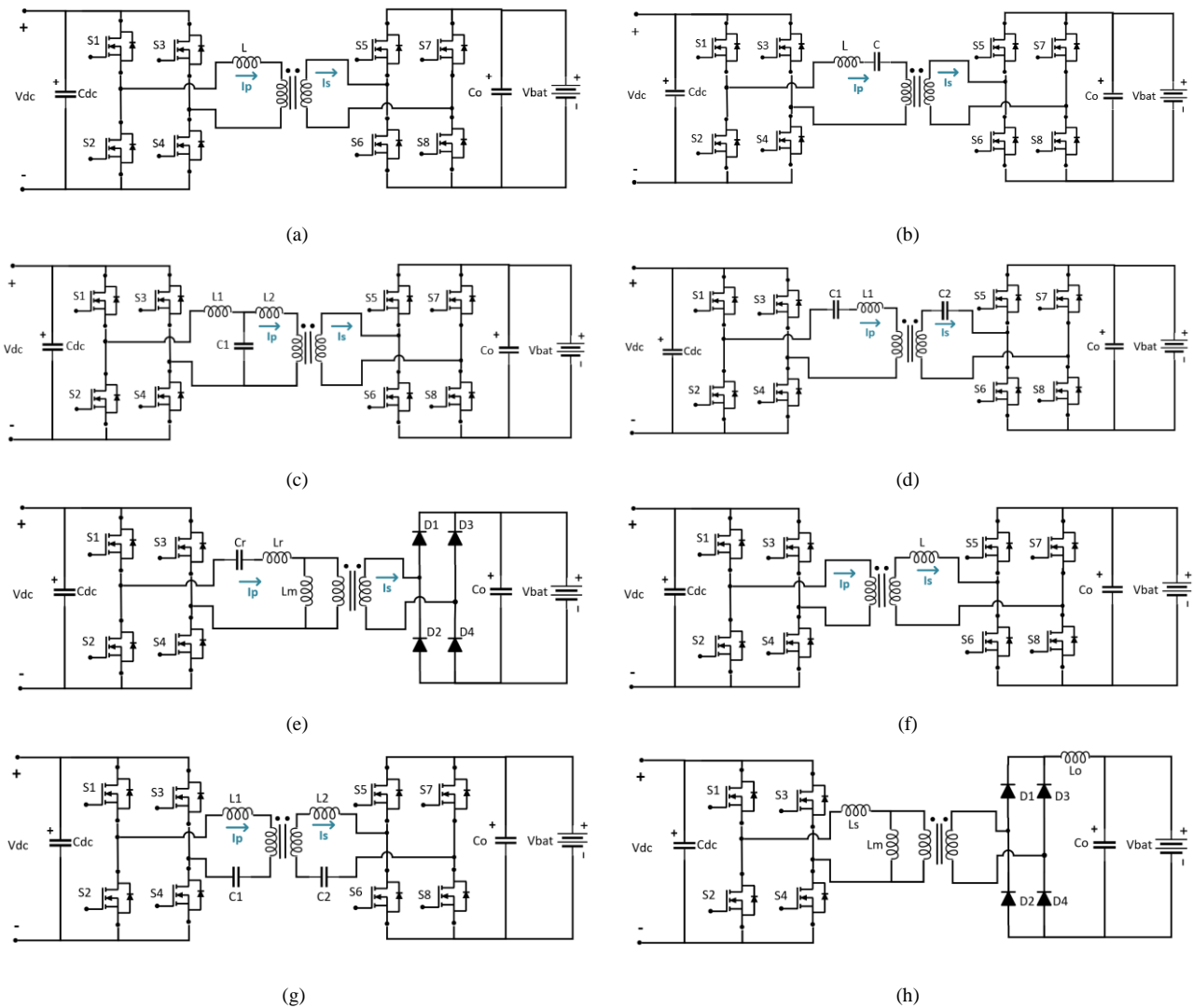

**FIGURE 15.** Isolated DC-DC Converter Configurations (a) Dual active bridge converter, (b) DAB series resonant converter (LC) (c) LCL resonant tank (d) CLC resonant tank (e) LLC resonant converter (f) LLL resonant converter (g) CLLC resonant tank (h) Phase-shifted full-bridge converter

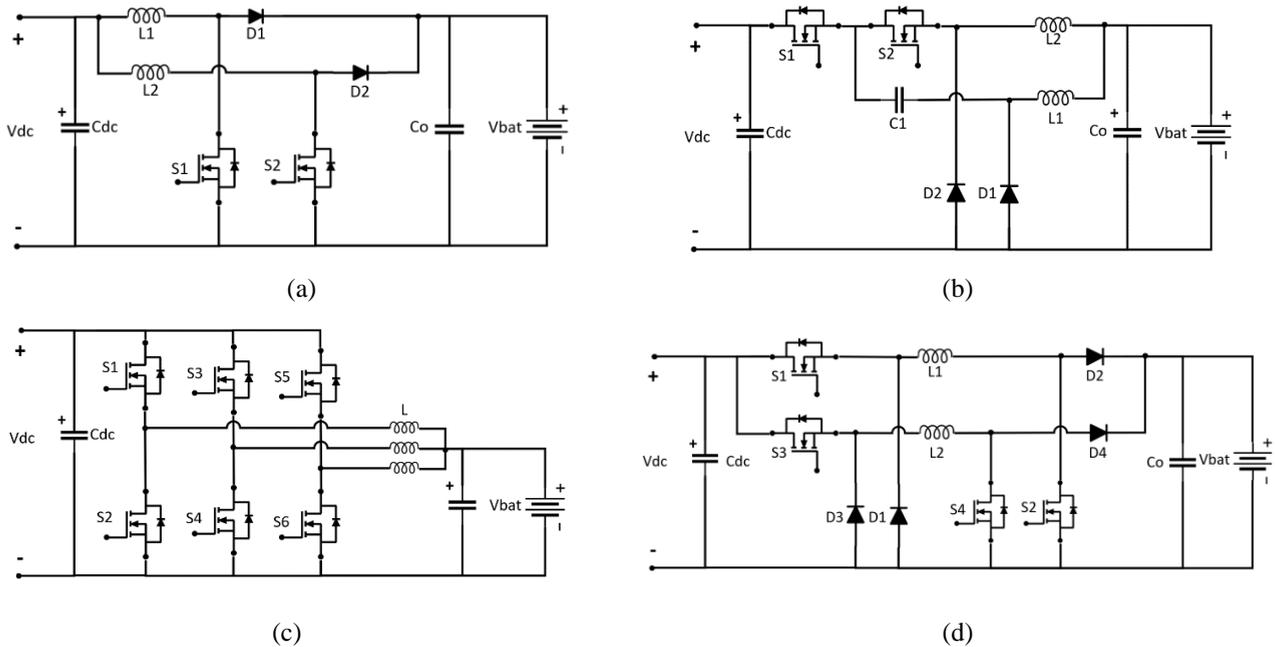

**FIGURE 16.** Non-Isolated DC-DC Converter Configurations (a) Interleaved boost converter (b) Interleaved two-phase buck converter (c) Interleaved Three-phase buck converter (e) Interleaved two-phase cascaded buck-boost converter

The resonant tank can be integrated between the primary and secondary bridges of the DAB converter with an improved modulation technique to extend the range of the ZVS of EV battery charging [198]. The research works have used various types of resonant tanks for DAB converters such as LC, LCL, CLC, LLL, and CLLC [199], [200], [201]. The series LC resonant tank with DAB is shown in Fig. 15(b). The LC resonant frequency is set less than the switching frequency of the converter and therefore it works in continuous current mode and may suffer from hard switching with a variation of battery voltages. The controlled switch inductor network is added in the DAB LC type converter to reduce the hard switching effects range and improve the operating battery voltage range [202]. The LCL and CLC type DAB resonant converters are shown in Fig. 15(c) and (d) respectively [203],[204]. The reactive power is considerably reduced as the bridge currents are in phase and sinusoidal and therefore the efficiency of these converters is very high. Moreover, conduction losses are very low compared to the DAB converter as they have a soft-switching range over all diodes and switchers. However, the use of a single central resonant tank may increase electric stress on the passive components and lead to an increase in current and voltage range [205], [206].

The LLC type DC-DC converter is widely used for EV charging systems due to its wide range of voltage output, ability to control voltage output at light load states, high efficiency, power density, and soft switching control capabilities [207]. The schematic diagram of DAB with LLC resonant tank is shown in Fig. 15(e) which is included a full-bridge converter, resonant tank, and filter capacitor. In [208], the two-stage onboard charger is designed with interleaved boost converter with a full bridge multi-resonant LLC converter. The efficiency of the proposed chargers is 95.4% and THD is less than 4% with an extendable voltage range. The current flow in the additional auxiliary inductor helps to achieve ZVS in LLL type DAB converter and the energy of the auxiliary inductor helps to charge and discharge the switching capacitance. The schematic diagram of the LLL tank-based DAB converter is shown in Fig. 15(f). The analysis of triple phase shift LLL tank-based DAB is presented in [198] which is a modified resonant tank with the controller to ensure efficient operation in various load conditions and voltage gains. Fig. 15(g) shows the schematic diagram of the DAB bidirectional CLLC resonant converter. The CLLC resonant converter has a wide range of output voltage, great soft switching, and symmetrical bidirectional power flow capabilities. However, the converter has low efficiency and voltage regulation challenges under light load conditions. The extended phase shift control scheme is introduced in [209] for CLLC type converter to improve efficiency in light load conditions.

The phase-shifted full bridge (PSFB) converters are commonly used for onboard chargers due to many advantages such as low EMI, current stress on the components, soft switching capability, simple PWM control, and high-power density [210]. In the schematic diagram of the PSFB converter shown in Fig. 15(h) that the secondary side active switches are replaced with the diodes for unidirectional power flow. However, the PSFB converters have conduction loss due to current circulation during the freewheeling gap, voltage overshoots across the secondary side, limited ZVS range, and duty cycle loss [211]. In [212], LLC resonant-based PSFB converter is presented to overcome the above challenges by using a hybrid modulation technique for LLC resonant converter. The PSFB converter with ZVS is presented in [213] to overcome conduction losses and high circulating energy of the converter. The primary side of the converter is built with the passive auxiliary circuit in [210] to ensure ZVS is activated for total battery range and secondary side voltage overshoots are decreased. Furthermore, a hybrid converter [214] and center tap clamp circuit [215] are integrated into the LLC type PDFB converter for the EV battery charging system to improve efficiency and power density.

### 2) NON-ISOLATED DC-DC CONVERTER

The non-isolated DC-DC converters are widely used in EV charging systems as they are simple, magnetic isolation free, low cost, light weight, high energy, and power density converters compared to the isolated DC-DC converters. They are compatible with line frequency transformer-based EV charging systems [216],[217]. The load and input of the non-isolated DC-DC converters share a common ground and in contrast, isolated converters have electrically isolated between the input and load [191]. The fundamental unidirectional non-isolated DC-DC converters are a buck, boost, buck-boost, SEPIC/Zeta, and Cuk and improved converter topologies are cascaded, interleaved, multilevel, bridge less, and modified bridge converters which are designed using basic converters. The boost converter is commonly used for the EV battery interface as the output voltage of the AC-DC converter is higher than the EV battery voltage. The interleaved boost converter is shown in Fig. 16 (a). The interleaved boost converters can provide better performances by reducing input/output ripples, conduction losses, high power, reduced size, and EMI in EV energy storage systems [218]. In [219], the boost PFC converter is presented with soft switching techniques to minimize the heating problem in the diode bridge and losses in boost diodes.

The single-phase buck converters can be employed for fast charging systems because the voltage of the EV battery is less than the input voltage of the DC-DC converter. The power rating is limited as one switch is preserving the total current and the inductor size is selected to be sufficiently

### TABLE XI
### COMPARISON OF ISOLATED AND NON-ISOLATED DC-DC CONVERTERS

| | Isolated Converter | Non-Isolated Converter |
|---|---|---|
| **Advantages** | - The low ripple of the output<br>- Advanced output regulation capabilities<br>- High performance of PF | - Simple design and control strategy<br>- Low cost and light weight<br>- High power and energy density<br>- Small input/output filter |
| **Disadvantages** | - Complex control and design<br>- Need additional components<br>- Expensive & increasing space and weight<br>- Low efficiency under light load | - The high voltage stress on switching devices<br>- High ripples in current and line frequency<br>- Discontinuous input current |

large to reduce charging/discharging losses caused by current ripples in the buck converter [220]. Therefore, interleaved buck converters with multiple inductors are used for EV charging systems to reduce inductor volume, current ripples, and heating impacts and increase power rating and efficiency [221]. A non-isolated interleaved two-phase and three-phase buck converters are shown in Fig. 16(b) and (c) respectively. The EV is designed with different battery voltage levels and therefore buck-boost converters are appropriate topologies to offer variable input/output voltages and bidirectional power flow in EV charging applications. The buck-boost converters can form in single or two stage power converters for charging applications with wide range of output voltage operation, high efficiency, high PFC performance, and voltage matching capabilities. The non-isolated onboard charger with the interleaved two-phase cascaded buck-boost converter is implemented in [222]. The proposed converter is shown in Fig. 16 (d) which can be achieved high power density, and efficiency with 97.6% and 0.99 PF in 3.7 kW of hardware implementation. A comparison of isolated and non-isolated DC-DC converts is presented in Table 11.

## VII. ELECTRIC VEHICLE BATTERY TECHNOLOGY

The EVs represent the largest share of the global battery market which is expected to the continuous growth of EV battery technologies, energy density, fast charging capabilities with long cycle life, and compliance with safety and environment standards [223], [224]. The battery is a key component of an EV that is capable to handle high energy capacity (kWh), and high power (kW) within limited weight, and space at an affordable price [225]. The EV battery is the main component in the EV charging system which is used to store electrical energy in the form of chemical energy when charging (G2V) and regenerative braking and feeds back to the power grid when discharging (V2G). Moreover, EV battery power is used to operate auxiliary elements including lights, and a cooling system. The advancement of battery technology drives the development of EVs as driving range and EV cost are determined by the battery energy density. The EV battery is connected to the DC-link via a DC-DC converter and the state of charge (SoC) demonstrates the control mechanisms of the battery. The type of battery is decided on the power requirement of the EV and various types of EV batteries are available in the market. Types of EV batteries are shown in Fig 17 and three main types of EV batteries are lead acid, nickel, and lithium-based batteries [226].

Lead-acid batteries are inexpensive, reliable, safe, and employed for high-power applications. However, they have low specific energy, a short lifetime, and weak performance in cold temperatures [227]. The main challenges in Ni-based batteries are high self-discharging, cost, heat generation at high temperatures, and required additional control systems to reduce losses. Lithium-ion (Li-ion) batteries are the dominant power storage in EVs due to their improved performances, high energy efficiency, energy storage, low self-discharge rate, and great performance in high temperatures. The Li-ion battery can charge and recharge regularly at any SoC. The Tesla cars have high power density NCA battery capacity (Tesla Model 3 has 80.5 kWh) and most of the EV manufacturers use LMO batteries due to their high specific power and energy. The total capacity of 62 kWh NCM is used in Nissan Leaf and the new generation Chevrolet Bolt EV (2020) consists of 68 kWh total battery capacity [228]. In [229], an extensive study of various control schemes for battery performance are evaluated under various situations including stability, multi-power resources, distributed network, and the different size of ESS in EV. Characteristics and specifications of commonly used EV battery types are presented in Table 12.

The modeling and SoC estimation of the EV battery is critical to optimize the energy management, and safety of the EV and extend battery life while reducing the cost of the vehicle. One of the key components of a battery managements ystem (BMS) is precise SoC estimation because it reflects battery performance and delivers essential factors for energy management and optimal control design for the vehicle [230], [231]. The accurate estimation of battery states is complex due to its non-linear and high time-varying behaviors. The SoC estimation is a key method used to maintain battery performance, which displays the remaining capacity of the battery via advanced algorithms with measurements. The current of each cell is remaining the same and an imbalance has occurred in cell voltage and SoC in a series of connected battery cells as the cell voltage and SoC of the battery differ from each other [232]. The SoC is the proportion of the remaining capacity of the battery to its rated capacity under a specific discharge rate. The primary function of the SoC is to communicate between the vehicle and the instinctive battery state to avoid over charging and discharging the battery [233] [234]. Furthermore, it provides critical information on available power, and battery usage until the next recharge, and executes a control system to improve the performance and life of the battery [235]. Numerous techniques have been proposed for real-time SoC estimation which can be classified into five groups model-based estimation, lookup table based, coulomb counting, data-driven estimation method, and hybrid method [234].

Model-based SoC estimate techniques such as equivalent circuit models, electrochemical models, and electrochemical impedance model are frequently used in EV charging [236]. The model-based SoC estimation techniques are accurate and powerful due to the reliance on the deep analysis of electrical, chemical, and combination of both characteristics. The comprehensive review of SoC estimation methods is presented in [237] by highlighting algorithm/control design, advantages, disadvantages, and challenges to select appropriate SoC methods for EVs. The comparison of existing SoC estimation methods and robust SoC estimation techniques are proposed in [238] based on the non-linear model and experimentally verified the effectiveness. In [239], a novel adaptive Kalman filter algorithm is designed for the SoC estimation of Li-ion batteries used in EVs. The improved deep neural network approach has been used in [240] to implement a new SoC estimation method for Li-ion batteries in EV applications.

The energy management system (EMS) is critical to address the driving range, battery life, efficiency, and reliable operation of EVs [241]. The EVs require multiple electronic control units to control and regulate power flow between converters, batteries, and motors. The battery management system (BMS) is the most responsible in the EV which is responsible for energy management by ensuring reliability and safety of the EV. The BMSs include sensors, a power delivery unit, and communication protocols to reduce the stress of the battery charging and discharging and prevent sudden abruption current to avoid high discharging rates. Moreover, cell balancing, calculating the SoC, computing the driving range, and other auxiliary are powered via the BMS. Smart coordinating with maximum RES can reduce the power load of the charging system on the grid and ensure cleaner energy. The V2G concept would be critical in improving the efficiency, overall quality, reliability, flexibility, and cost-effectiveness of the utility system [242].

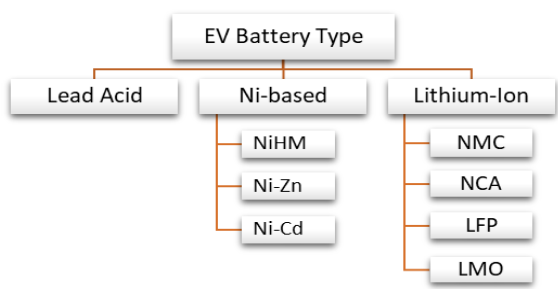

**FIGURE 17.** Types of Electric Vehicle Battery

| Battery Type | Vehicle Model | Specific energy (Wh/kg) | Energy density (Wh/L) | Cycle life | Safety | Specifications |
|---|---|---|---|---|---|---|
| Lithium Nickel Cobalt Aluminum Oxide (NCA) | Tesla X, S, 3, Y | 200-260 | 600 | 500 | Good | • Provide good energy yield and is inexpensive.<br>• Extensively used in both portable electronics and EVs |
| Lithium Nickel Manganese Cobalt Oxide (NMC) | Nissan Leaf, Kia e-Soul, Volkswagen e-Golf, BMW i3, I3s Peugeot e-208 | 150-220 | 580 | 1000-2000 | Good | • Stable chemistry, and relatively low-cost materials<br>• Provide a high energy density and able to charge rapidly compared to other batteries. |
| Lithium Manganese Oxide (LMO) | Chevy-Volt, Escape PHEV | 100-150 | 420 | 300-700 | Good | • Good energy performance and low cost of materials.<br>• Short life cycle. |
| Lithium Iron Phosphate (LFP) | EVs, especially in e-bikes, e-rikshaw, | 90-120 | 330 | 1000-2000 | Excellent | • Stable, long lifecycle, and significant safety.<br>• High energy density and low rate of self-discharge make it ideal for larger EVs such as vans, buses, or trucks. |
| Lithium Titanate (LTO) | Mitsubishi, Honda | 50-80 | 130 | 3000-7000 | Excellent | • Long life, fast charge using advanced Nanotechnology<br>• very high rate of charging and discharging possible without compromising on safety |

## VIII. ARCHITECTURES OF ELECTRIC VEHICLE CHARGING STATION

The charging infrastructure of EVs is main deployed to meet the user requirements and provide convenient and reliable charging and discharging of the EV battery. The EV charging station can perform similar to a fuel station by supplying direct current from the grid, renewable energy source, or ESS with the ability to rapidly charge. The fast-charging stations are connected to the medium voltage network to supply high power from the grid. Therefore, they required high capital investments to design additional control techniques to maintain power requirements and standards on both sides of the fast-charging station. Renewable energy sources and energy storage systems are widely preferred in the present EV charging station to minimize impacts on the grid while providing additional network services. Moreover, charging stations with V2G capabilities are widely researched the present to enhance grid support. The architecture of EV charging stations can be classified as AC bus, DC bus, and a combination of AC and DC bus structures based on the power supply requirement.

### A. CONVENTIONAL CHARGING STATIONS

The three-phase AC bus operated between the 250V-480V line-to-line voltage in the AC bus-connected fast charging stations [34]. The charging unit in the station consists of a DC-DC converter and AC-DC rectifier which causes an increase in power conversion stages, complexity, cost, and reduced efficiency of the system. The secondary side of the step-down transformer serves as a common AC bus in the architecture shown in Fig. 18(a), and individual

AC-DC converters and DC-DC converters are employed to link each EV load to the common AC bus. In contrast, DC bus-connected charging units use a single AC-DC converter to provide common DC bus services and each EV load is employed with an independent DC-DC converter as shown in Fig. 18(b). Hence, DC-bus architectures are more efficient, cost-effective, small, and more flexible structures with greater dynamic performance when compared to the AC bus-based architecture. The DC bus system also offers a more adaptable structure with the possibility to connect ESS and RES. However, low operating PF of DC-bus structures EV charging stations can generate undesirable harmonic impacts on the utility grid.

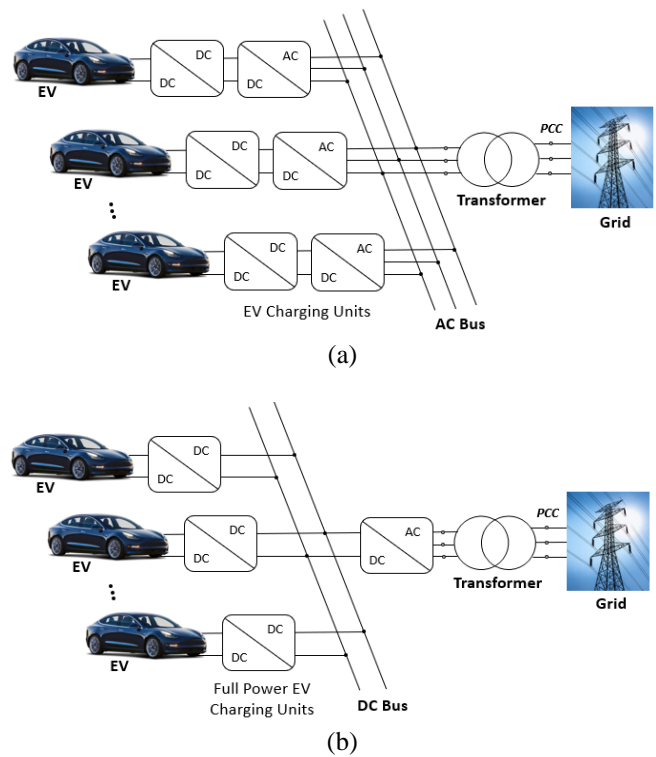

**FIGURE 18.** Architecture of conventional EV charging station (a) Common AC bus-based (b) Common DC bus-based system

AC charging stations are preferred as public charging stations due to their low manufacturing cost and consist of matured AC technology and standard charging components in the market. The slow charging application in AC bus systems has a maximum of 19.2 kW power [246]. The AC bus connected fast, and ultra-fast charging stations are necessary to be equipped with advanced components and controllers to maintain grid codes and EV charging standers. Therefore, DC bus architecture is commonly preferred for fast and ultra-fast EV charging stations as they have a low impact on the utility grid, a simple control strategy, and high efficiency. The DC and fast charging (22kW - 200 kW) and ultra-fast charging (>300 kW) capabilities are commonly designed in off-board chargers with high power flow and galvanic isolation is mandated between EV battery and the grid according to the IEC standards [247]. The comparative analysis of AC and DC bus architectures is presented in [248] for a grid-connected fact EV charging system. The power quality of the two systems is evaluated under dynamic and steady-state conditions and different transformer configurations and concluded that common DC bus architecture has better performance than AC bus architecture. The comparison of AC and DC bus-based charging systems is summarized in Table 13.

## B. AC AND DC BUS-BASED CHARGING STATIONS

The AC and DC bus-based architectures are generally considered as a DC grid and AC micro-grid which are particularly employed with DC power sources. This hybrid architecture of EV charging stations comprises the power grid and different energy sources which are connected to the AC and DC busses via separate converters as shown in Fig. 19. This configuration provides simultaneous operation of both AC and DC charging by preventing additional power conversion states [118]. A single bidirectional converter is employed to connect AC and DC buses in the system which is called an interlinked power converter and the corresponding buses can be used to connected AC and DC loads. The interlink converter can maintain energy balance between both sides and operate according to the load requirements. This architecture is very reliable, flexible, and more efficient than AC and DC bus configurations. The bidirectional DC-DC converter is connected between the EV and the DC bus to achieve fast DC charging and discharging via V2G operation. The AC and DC bus-based structure is used to investigate microgrids and ESS [249]. The stand-alone V2G control technique in proposed in [250] to examine charging and discharging performance and RES power characteristics in hybrid AC-DC charging architecture.

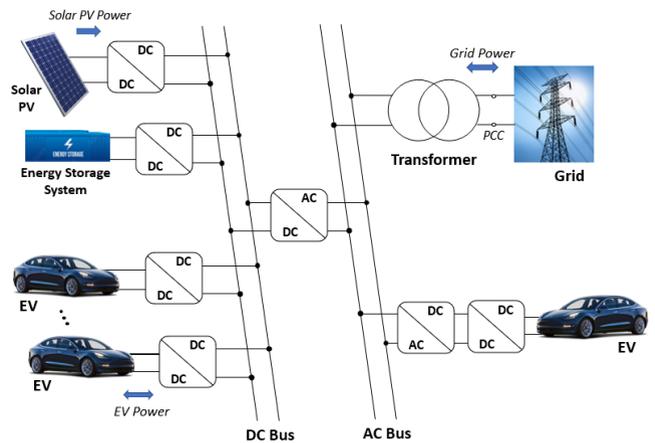

FIGURE 19. Architecture of AC and DC bus-based EV charging station

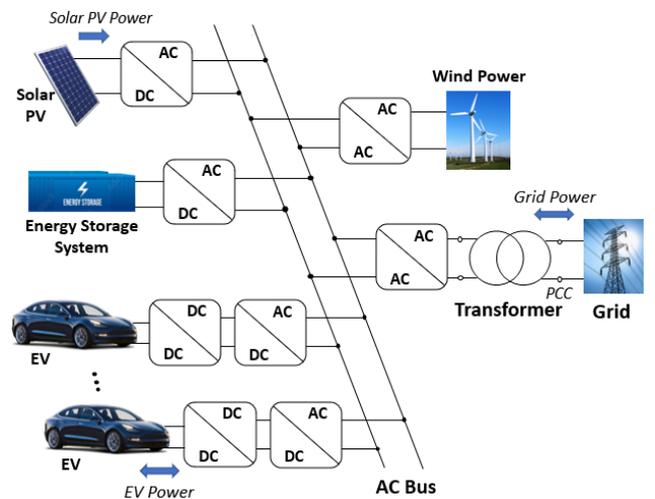

FIGURE 20. Architecture of renewable energy sources and energy storage system connected common AC bus -based EV charging station

## C. Renewable Energy Integrated Charging Station

The EV charging station may be supported by the power grid, standalone RESs, or combined grid-connected RESs depending on the power grid's availability to prevent local power network overload and ensure a higher proportion of clean energy usage. Most researchers attempt to enable high renewable energy-based power generation on the EV charging stations to decrease the power demand during the charging period by managing their charging patterns. The architecture RES connected common AC- bus-based EV charging station is shown in Fig. 20 including the power grid, solar PV, wind power, and ESS and EV chargers with relevant converters and control units. The AC bus of the above (Fig. 20) architecture can be changed by using a common DC bus with reduced converter stages. RES-based charging systems are increasingly developed due to various factors. The implementation of RESs and EVs deliver an exceptional opportunity for sustainable charging of EVs which can be directly utilized to charge EVs during peak time [251].

TABLE XIII
COMPARISON OF AC AND DC BUS-BASED CHARGING STATIONS ARCHITECTURES

| Architecture | Advantages | Disadvantages |
|---|---|---|
| AC bus-based EV charging systems | • Highly availability and mature technology with standards.<br>• The complexity of the protection devices is low.<br>• Able to direct usage for local loads<br>• Stability and scalability are high<br>• Reliable switching and control techniques<br>• Ability to control active and reactive power | • A large number of converters reduce rated power and efficiency.<br>• High cost due to multiple converters.<br>• Additional conversion stages are required for fast chargers to avoid harmonics.<br>• Difficult to achieve high power quality and stability<br>• Complex to integrate RESs and required additional DC/DC stage. |
| DC bus-based EV charging systems | • Provide high efficiency and power density due to low components.<br>• The control strategy is simple<br>• Low cost and flexible configuration which can integrate ESS and RES easily.<br>• Helps to reduce the impacts of high penetration of EV loads on the power grid<br>• Reduction in frequency fluctuation | • Complexity may increase with the additional energy sources.<br>• Required protection devices to withstand sudden changes<br>• The central converter needs server conditions due to an increase in nominal power. |

Solar PV integrated EV charging systems can be employed to reduce peak demand by decreasing EV reliance on grid power. Solar PV panels have been installed rapidly, are more affordable at low cost, and EV batteries can be used to store energy for solar PV as they can consume a large amount of Solar PV energy [252], [253]. Many researchers have discovered that coordinate operation of PVs and EVs can decrease impacts encountered by individual PV and EV integration on the power grid [254]. However, the integration of EVs and RESs into the grid is a challenge due to the additional planning stages, converters, and control strategies are needed to be considered in this type of charging station. The uncontrolled or uncoordinated effect on system consistency can be compromised and introduced many negative impacts on the distribution grid [255],[256]. The RESs connected to common DC bus-based EV charging architecture are extensively researched over other structures due to efficiency and flexibility to integrate different energy resources, and smart control capabilities [118].

## IX. FUTURE TRENDS AND CHALLENGES

The exponential growth of EVs over the past decade has created an advanced electrified transportation system as well as new challenges on the distribution grid. Manufactures are attempting to overcome driving range limitations, higher upfront costs, longer recharge time, EV battery-related constraints, and limited charging facilities [257]. Therefore, Evaluation of the emerging technologies, control strategies, and future trends in EV charging systems is important for exploring the value of finding new solutions and improvements. EV will reach a power density of 33 kW/L, 480 000 km/ 15-year lifetime, and 100 kW electric drive capacity in 2025 according to the Department of Energy roadmap [258]. Automated EVs are one of the trending topics in the BEV industry due to design freedom in the same structure. Therefore, predictions indicate that performance and customer satisfaction will further increase while reducing the operational and charging costs of EVs. Ultra-fast charging technologies also increasingly developing to make sure the same experience as ICE vehicles at gas stations.

Wireless charging will likely acquire market acceptance due to many advantages including extreme fast charging capabilities, low maintenance, reduced components (ports, connectors, and cables), and automated charging capabilities. Wireless charging systems are considered as safe, cost-effective flexible, and reliable charging method. The resonant wireless power transfer technique is most preferred and future challenges and opportunities of wireless charging systems are reviewed in [259]. V2G and V2X (V2H, V2B) technologies are developed to enable feed EV battery energy back according to the user requirements [260]. V2G operation has smart charging control which can balance variations in energy consumption and production. V2G solutions are ready to reach the market and enhance grid support. The intelligent algorithm designed is another accepted revolution of future EV technology to further improve electrical infrastructure. Intelligent methods are commonly employed in BMS to address dynamic, complex, and non-linear attributes of EV batteries [261], [262]. Moreover, they can predict future states based on previous information and improve the performance of the charging system.

Internet of Vehicles (IoV) is one of the most promising applications in developing intelligent transportation systems in smart cities. The network of the EV is equipped with communications technology to improve transportation infrastructure by providing parking assistance, efficient vehicle maintenance, energy saving, and reducing traffic congestion [263], [264]. Cost and size are key challenges to achieving high power density in EV charging systems. The cost and size of the EV are mainly depending on the energy storage system. The improvements in battery charging density, charging and discharging methods, materials, durability, and SoC estimation method are critical factors of energy storage systems as they determine the cost and performance of the EV. Automakers are attempting to develop long-rage high power density EVs with limited space. Replacing silicon switches with wide bandgap power electronic devices is considered to reduce the size of the power modules, and the high temperature and frequency impacts of the charging system [258].

Lack of the EV charging stations could limit EV growth and range anxiety is increased as not enough charging stations as well as long charging time. The recharging time of the EV is higher than the refueling time of an ICE vehicle. The maximum charging rate of EV and charging point, battery size, and SoC estimation method are the main factors associated with battery charging time. Local regulatory issues and electric grid upgrades also play an important role in growing charging stations [265]. Adaptive charging strategies can be used in public parking lots to limit simultaneous power in charging points. Moreover, the integration of intelligent load balancing systems may help to manage efficient charging systems, and grid performances and satisfy the charging demands. The high level of EV integration into the distribution network has introduced many technical challenges for power grid operations, safety, and network planning due to increased load demand, voltage and current fluctuations, power quality impacts, and power losses [266]. Modern EV charging systems are tended to use other energy sources link solar PV, wind power, and energy storage systems. Therefore, the intermittency nature and grid integration of RESs also need to be controlled to minimize grid impacts [120]. Various control strategies have been employed to improve charging/discharging capabilities and mitigate impacts on the distribution grid [267],[268].

## X. CONCLUSION

The large-scale adoption of electric vehicles necessitates investigate and development of charging technologies and power converters to achieve high efficient, low-cost, and reliable charging solutions for an EV battery. This paper has reviewed electric vehicle charging topologies, configurations, and architectures of charging stations. The status and requirements of EV charging systems are evaluated by considering standards, charging levels, modes, and EV supply equipment. Conductive charging topologies can be divided into onboard and offboard chargers, unidirectional and bidirectional as well as AC and DC chargers according to the emerging technology. Onboard and offboard charging are discussed comprehensively with examples to understand the powertrain of different chargers. Integrated chargers are gaining more attention due to their many advantages over dedicated onboard chargers. Integrated solutions can overcome the cost, weight, volume, and power limitations of conventional onboard chargers and can control charging voltage via drive inverter and motor inductance without employing separate stages. Modularity concept seems to play the main success toward ultra-fast and fast charging and enable versatility in different power electronic solutions. Power converter configurations of EV charging systems are presented based on AC-DC and DC-DC converters with circuit topologies and explained.

Standardization of charging requirements, infrastructure decisions, smart control strategies and enhanced battery technologies are essential to successful EVs adoption. The performance of the battery not only depends on design and type but also characteristics of the charger, charging and discharging infrastructure, and the SoC estimation method. Moreover, a comparative analysis has been carried out for the charging station architecture in terms of AC/DC power flow, control strategy, advantages, and disadvantages. Most multiport EV charging stations are integrated with solar PV and energy storage to reduce stress on the utility grid while providing ancillary services. Finally, future trends and challengers are evaluated in terms of the technical limitations, charging/discharging capabilities, smart charging, battery performance, and grid integration. This paper has provided a detail view of EV charging system technologies by highlighting the advances to motivate the development of novel designs.